\documentclass[review]{elsarticle}

\usepackage{tikz}
\usepackage{verbatim}
\usepackage{textcomp}
\usepackage{algpseudocode}
\usepackage{mdframed}
\usepackage[normalem]{ulem}
\usepackage{epstopdf}
\usepackage{graphicx}
\usepackage{subfig}
\usepackage{natbib}
\setcitestyle{numbers,sort&compress}
\usepackage{algorithm}
\usepackage{algpseudocode}
\usepackage{mwe}

\usetikzlibrary{backgrounds, calc,trees,positioning,arrows,chains,shapes.geometric, decorations.pathreplacing,decorations.pathmorphing,shapes,patterns, matrix,shapes.symbols,fit,scopes,external}

\usepackage{pgfplots,pgfplotstable}
\pgfplotsset{compat=newest}

\usepackage{booktabs} 
\usepackage{balance}
\usepackage{listings}
\usepackage{xcolor} 
\usepackage{soul}

\usepackage{amsmath,amssymb,amsfonts}
\usepackage{lipsum}
\usepackage{array,multirow}
\usepackage{siunitx}
\usepackage[acronyms,nonumberlist,nopostdot,nomain,nogroupskip]{glossaries}
\usepackage{lineno}

\usepackage{hyperref}

\modulolinenumbers[5]

\journal{Simulation Modelling Practice and Theory}

\definecolor{seablue}{RGB}{05,102,141}
\definecolor{alabamacrimson}{RGB}{172,00,54}
\definecolor{darkgreen}{RGB}{03,49,46}
\definecolor{metallicsunburst}{RGB}{168,118,62}
\definecolor{silver}{RGB}{193,193,193}

\long\def\/*#1*/{}

\newacronym{quic}{QUIC}{Quick UDP Internet Connections}
\newacronym{3gpp}{3GPP}{3rd Generation Partnership Project}
\newacronym{adc}{ADC}{Analog to Digital Converter}
\newacronym{5g}{5G}{5th generation}
\newacronym{aimd}{AIMD}{Additive Increase Multiplicative Decrease}
\newacronym{am}{AM}{Acknowledged Mode}
\newacronym{amc}{AMC}{Adaptive Modulation and Coding}
\newacronym{aqm}{AQM}{Active Queue Management}
\newacronym{awgn}{AGWN}{Additive White Gaussian Noise}
\newacronym{balia}{BALIA}{Balanced Link Adaptation}
\newacronym{bdp}{BDP}{Bandwidth-Delay Product}
\newacronym{bf}{BF}{Beamforming}
\newacronym{cc}{CC}{Congestion Control}
\newacronym{cdf}{CDF}{Cumulative Distribution Function}
\newacronym{cn}{CN}{Core Network}
\newacronym{cqi}{CQI}{Channel Quality Information}
\newacronym{cp}{CP}{Control Plane}
\newacronym{csirs}{CSI-RS}{Channel State Information - Reference Signal}
\newacronym{dc}{DC}{Dual Connectivity}
\newacronym{dce}{DCE}{Direct Code Execution}
\newacronym{dci}{DCI}{Downlink Control Information}
\newacronym{dl}{DL}{Downlink}
\newacronym{dmr}{DMR}{Deadline Miss Ratio}
\newacronym{dmrs}{DMRS}{DeModulation Reference Signal}
\newacronym{e2e}{E2E}{End-to-End}
\newacronym{ecn}{ECN}{Explicit Congestion Notification}
\newacronym{edf}{EDF}{Earliest Deadline First}
\newacronym{enb}{eNB}{evolved Node Base}
\newacronym{epc}{EPC}{Evolved Packet Core}
\newacronym{es}{ES}{Edge Server}
\newacronym{fdma}{FDMA}{Frequency Division Multiple Access}
\newacronym{fdd}{FDD}{Frequency Division Duplexing}
\newacronym[firstplural=Radio Access Technologies (RATs)]{rat}{RAT}{Radio Access Technology}
\newacronym{fs}{FS}{Fast Switching}
\newacronym{ftp}{FTP}{File Transfer Protocol}
\newacronym{gnb}{gNB}{Next Generation Node Base}
\newacronym{harq}{HARQ}{Hybrid Automatic Repeat reQuest}
\newacronym{hetnet}{HetNet}{Heterogeneous Network}
\newacronym{hh}{HH}{Hard Handover}
\newacronym{hol}{HOL}{Head-of-Line}
\newacronym{ia}{IA}{Initial Access}
\newacronym{imt}{IMT}{International Mobile Telecommunication}
\newacronym{iot}{IoT}{Internet of Things}
\newacronym{ipsd}{IPSD}{Interference Power Spectral Density}
\newacronym{kpi}{KPI}{Key Performance Indicator}
\newacronym{kpis}{KPIs}{Key Performance Indicators}
\newacronym{los}{LOS}{Line of Sight}
\newacronym{lte}{LTE}{Long Term Evolution}
\newacronym{m2m}{M2M}{Machine to Machine}
\newacronym{mac}{MAC}{Medium Access Control}
\newacronym{mc}{MC}{Multi-Connectivity}
\newacronym{mcs}{MCS}{Modulation and Coding Scheme}
\newacronym{mec}{MEC}{Mobile Edge Cloud}
\newacronym{mi}{MI}{Mutual Information}
\newacronym{mimo}{MIMO}{Multiple-Input Multiple-Output}
\newacronym{mmwave}{mmWave}{millimeter wave}
\newacronym{mr}{MR}{Maximum Rate}
\newacronym{mss}{MSS}{Maximum Segment Size}
\newacronym{mtd}{MTD}{Machine-Type Device}
\newacronym{mtu}{MTU}{Maximum Transmission Unit}
\newacronym{nfv}{NFV}{Network Function Virtualization}
\newacronym{nlos}{NLOS}{Non Line of Sight}
\newacronym{nr}{NR}{New Radio}
\newacronym{ofdm}{OFDM}{Orthogonal Frequency Division Multiplexing}
\newacronym{o2i}{O2I}{Outdoor-To-Indoor}
\newacronym{pdcch}{PDCCH}{Physical Downlonk Control Channel}
\newacronym{pdcp}{PDCP}{Packet Data Convergence Protocol}
\newacronym{pdsch}{PDSCH}{Physical Downlink Shared Channel}
\newacronym{pdu}{PDU}{Packet Data Unit}
\newacronym{pf}{PF}{Proportional Fair}
\newacronym{pgw}{PGW}{Packet Gateway}
\newacronym{phy}{PHY}{Physical}
\newacronym{pbch}{PBCH}{Physical Broadcast Channel}
\newacronym[plural=\gls{mme}s,firstplural=Mobility Management Entities (MMEs)]{mme}{MME}{Mobility Management Entity}
\newacronym{prb}{PRB}{Physical Resource Block}
\newacronym{pss}{PSS}{Primary Synchronization Signal}
\newacronym{pucch}{PUCCH}{Physical Uplink Control Channel}
\newacronym{pusch}{PUSCH}{Physical Uplink Shared Channel}
\newacronym{rach}{RACH}{Random Access Channel}
\newacronym{ran}{RAN}{Radio Access Network}
\newacronym{rbg}{RBG}{Resource Block Group}
\newacronym{red}{RED}{Random Early Detection}
\newacronym{rem}{REM}{Radio Environment Map}
\newacronym{rems}{REMs}{Radio Environment Maps}
\newacronym{rf}{RF}{Radio Frequency}
\newacronym{rlc}{RLC}{Radio Link Control}
\newacronym{rlf}{RLF}{Radio Link Failure}
\newacronym{rrc}{RRC}{Radio Resource Control}
\newacronym{rrm}{RRM}{Radio Resource Management}
\newacronym{rr}{RR}{Round Robin}
\newacronym{rs}{RS}{Remote Server}
\newacronym{rsrp}{RSRP}{Reference Signal Received Power}
\newacronym{rsrq}{RSRQ}{Reference Signal Received Quality}
\newacronym{rssi}{RSSI}{Received Signal Strength Indicator}
\newacronym{rss}{RSS}{Received Signal Strength}
\newacronym{rtt}{RTT}{Round Trip Time}
\newacronym{rv}{RV}{Reduncy Version}
\newacronym{rw}{RW}{Receive Window}
\newacronym{rx}{RX}{Receiver}
\newacronym{sa}{SA}{standalone}
\newacronym{sack}{SACK}{Selective Acknowledgment}
\newacronym{sap}{SAP}{Service Access Point}
\newacronym{sch}{SCH}{Secondary Cell Handover}
\newacronym{scm}{SCM}{Spatial Channel Model}
\newacronym{scoot}{SCOOT}{Split Cycle Offset Optimization Technique}
\newacronym{sdma}{SDMA}{Spatial Division Multiple Access}
\newacronym{sinr}{SINR}{Signal to Interference plus Noise Ratio}
\newacronym{sm}{SM}{Saturation Mode}
\newacronym{snr}{SNR}{Signal to Noise Ratio}
\newacronym{son}{SON}{Self-Organizing Network}
\newacronym{ss}{SS}{Synchronization Signal}
\newacronym{srs}{SRS}{Sounding Reference Signal}
\newacronym{sss}{SSS}{Secondary Synchronization Signal}
\newacronym{tb}{TB}{Transport Block}
\newacronym{tcp}{TCP}{Transmission Control Protocol}
\newacronym{tdd}{TDD}{Time Division Duplexing}
\newacronym{tdma}{TDMA}{Time Division Multiple Access}
\newacronym{tfl}{TfL}{Transport for London}
\newacronym{tm}{TM}{Transparent Mode}
\newacronym{trp}{TRP}{Transmitter Receiver Pair}
\newacronym{tti}{TTI}{Transmission Time Interval}
\newacronym{ttt}{TTT}{Time-to-Trigger}
\newacronym{tx}{TX}{Transmitter}
\newacronym{ue}{UE}{User Equipment}
\newacronym{ul}{UL}{Uplink}
\newacronym{uml}{UML}{Unified Modeling Language}
\newacronym{um}{UM}{Unacknowledged Mode}
\newacronym{utc}{UTC}{Urban Traffic Control}
\newacronym{vm}{VM}{Virtual Machine}
\newacronym{crs}{CRS}{Cell Reference Signal}
\newacronym{comp}{CoMP}{Coordinated Multi-Point}
\newacronym{cran}{C-RAN}{Cloud \acrlong{ran}}
\newacronym{ca}{CA}{Carrier Aggregation}
\newacronym{cco}{CC}{Carrier Component}
\newacronym{nsa}{NSA}{Non Stand Alone}
\newacronym{embb}{eMBB}{Enhanced Mobility Broadband}
\newacronym{bsr}{BSR}{Buffer Status Report}
\newacronym{srb}{SRB}{Service Radio Bearer}
\newacronym{sctp}{SCTP}{Stream Control Transmission Protocol}
\newacronym{mptcp}{MPTCP}{Multi-path TCP}
\newacronym{ietf}{IETF}{Internet Engineering Task Force}
\newacronym{os}{OS}{Operating System}
\newacronym{tls}{TLS}{Transport Layer Security}
\newacronym{rfc}{RFC}{Request for Comments}
\newacronym{http}{HTTP}{HyperText Transfer Protocol}
\newacronym{nat}{NAT}{Network Address Translation}
\newacronym{api}{API}{Application Programming Interface}
\newacronym{rto}{RTO}{Retransmission Timeout}
\newacronym{psc}{PSC}{Public Safety Communication}
\newacronym{rpgm}{RPGM}{Reference Point Group Mobility}
\newacronym{ic}{IC}{Incident Command}
\newacronym{rsu}{RSU}{Road Side Unit}
\newacronym{uav}{UAV}{Unmanned Aerial Vehicle}
\newacronym{iab}{IAB}{Integrated Access and Backhaul}
\newacronym{psd}{PSD}{Power Spectral Density}
\newacronym{mpc}{MPC}{Multi Path Component}
\newacronym{rt}{RT}{Ray Tracer}
\newacronym{aoa}{AoA}{Angle of Arrival}
\newacronym{aod}{AoD}{Angle of Departure}
\newacronym{inr}{INR}{Interference to Noise Ratio}
\newacronym{qd}{QD}{Quasi Deterministic}
\newacronym{wlan}{WLAN}{Wireless Local Area Network}
\newacronym{cad}{CAD}{Computer-aided Design}
\newacronym{ap}{AP}{Access Point}
\newacronym{sta}{STA}{Station}
\newacronym{nrmse}{NRMSE}{Normalized Root Mean Square Error}
\newacronym{ut}{UT}{User Terminal}
\newacronym{bs}{BS}{Base Station}

\begin{document}
\begin{frontmatter}

\title{MIMO in network simulators: Design, implementation and evaluation of single-user MIMO in ns-3 5G-LENA}

\author{Biljana Bojovi\'c}
\ead{biljana.bojovic@cttc.es}

\author{Sandra Lag\'en}
\ead{sandra.lagen@cttc.es}

\address{Centre Tecnol\`ogic de Telecomunicacions de Catalunya (CTTC/CERCA), 
Avinguda Carl Friedrich Gauss, 7, 08860 Castelldefels, Barcelona, Spain 
}
 
\begin{abstract}
MIMO technology has been studied in textbooks for several decades, and it has been adopted in 4G and 5G systems. Due to the recent evolution in 5G and beyond networks, designed to cover a wide range of use cases with every time more complex applications, it is essential to have network simulation tools (such as ns-3) to evaluate MIMO performance from the network perspective, before real implementation. Up to date, the well-known ns-3 simulator has been missing the inclusion of single-user MIMO (SU-MIMO) models for 5G. In this paper, we detail the implementation models and provide an exhaustive evaluation of SU-MIMO in the 5G-LENA module of ns-3. As per 3GPP 5G, we adopt a hybrid beamforming architecture and a closed-loop MIMO mechanism and follow all 3GPP specifications for MIMO implementation, including channel state information feedback with precoding matrix indicator and rank indicator reports, and codebook-based precoding following Precoding Type-I (used for SU-MIMO). The simulation models are released in open-source and currently support up to 32 antenna ports and 4 streams per user. The simulation results presented in this paper help in testing and verifying the simulated models, for different multi-antenna array and antenna ports configurations.
\end{abstract}


\end{frontmatter}

\section{Introduction}
The use of multiple antennas in communication systems (a.k.a., \gls{mimo} systems) has attracted a lot of attention in the past decades. MIMO permits increasing the data rate by sending multiple data streams simultaneously (known as spatial multiplexing) or increasing the robustness of the data transmission by sending replicated data (known as transmit diversity) or providing array gain into specific spatial areas by properly designing the antenna weights or beamforming vectors (known as beamforming)~\cite{842121,palomar:phd}. Beamforming is particularly essential at high carrier frequencies within the \gls{mmwave} region to combat the high pathloss propagation losses and blocking effects~\cite{pi:11,andrews:17,lagen:19}. On the other hand, spatial multiplexing has the ability to approach the maximum channel capacity, and at the same time adapt the number of transmitted data streams (or layers) to the available propagation conditions. Due to the various trade-offs between beamforming and spatial multiplexing, in terms of performance and implementation complexity, hybrid precoding architectures arose as a very promising MIMO solution for 5G New Radio (NR) and future mobile network systems~\cite{TS38214}. In hybrid precoding architectures, digital baseband processing (for spatial multiplexing plus digital beamforming, also known as precoding) and analog processing (for analog beamforming) are combined~\cite{8371237}.

\begin{figure}[!t]
\includegraphics[width=0.7\columnwidth]{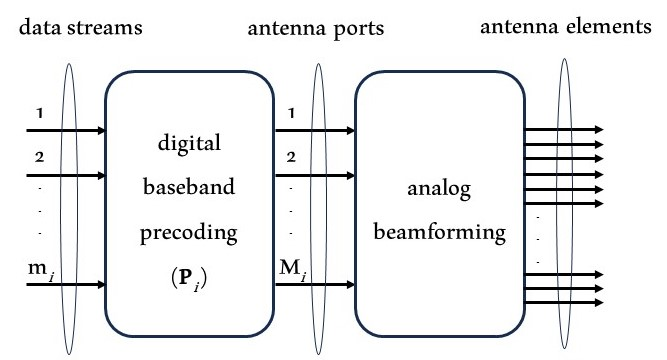}
\centering
\caption{Hybrid precoding structure for MIMO systems in 3GPP 5G NR.}
\label{fig:mimo}
\end{figure}

The 3rd Generation Partnership Project (3GPP) has already adopted the hybrid precoding architecture for MIMO in 5G NR, by introducing the concept of antenna ports, as shown in Fig.~\ref{fig:mimo}. This way, multiple data streams are mapped into multiple antenna ports by means of digital baseband processing (precoding), and then the multiple antenna ports are mapped to antenna elements by means of analog beamforming. 
For the antenna ports to antenna elements mapping, 3GPP supports either a sub-connected structure (each antenna port is connected to a unique subset of antenna elements) or a fully connected structure (each antenna port is connected to all antenna elements). 
For the data streams to antenna ports mapping, 3GPP has adopted for 5G NR a closed-loop MIMO mechanism with codebook-based precoding, to reduce the feedback complexity and overhead. In closed-loop MIMO, the Channel State Information (CSI) is acquired at the base station (gNB) thanks to CSI feedback from the user equipment (UE), which includes the Precoding Matrix Indicator (PMI), the Rank Indicator (RI) (i.e., the number of streams), and the Channel Quality Indicator (CQI)~\cite{TS38214}. Owing to the codebook-based precoding, the PMI selection (done at the UE side) results in an index of a set of predefined precoding matrices in a codebook. 
5G NR supports two types of codebooks for MIMO: Type-I and Type-II~\cite{TS38214}. Type-I codebook is mainly used for single-user MIMO (SU-MIMO) (which sends multiple streams towards the same user) and it is based on the same logic as LTE codebook. On the other hand, Type-II codebook is mostly for multi-user MIMO (MU-MIMO) (which sends multiple streams but each stream towards a different user) and it adopts a more general mathematical formula with many of parameters.

To support the evaluation of MIMO technology in 5G mobile networks, system-level simulators that include an end-to-end perspective of the network with a complete protocol stack but at the same time, accurate physical (PHY) layer modeling for MIMO are needed. 
Network simulators such as \texttt{ns-3}~\cite{ns3} offer a powerful tool to evaluate different ``hot topic'' 5G scenarios through complete network modeling and complex applications, such as augmented, virtual, and mixed reality in 5G and beyond~\cite{10120945,RP-201145}. At the same time, to allow the simulation of such demanding applications, it is necessary to enable realistic MIMO models in the simulation, including SU-MIMO and MU-MIMO. Accurate MIMO models are needed to be able to assess realistic XR application layer performance in 5G, which can be greatly affected by the performance of different MIMO implementations, i.e., 
PMI search/selection algorithms. In parallel, while on the road towards 6G, there is more and more interest in the research community in energy-efficient solutions for UE devices. Note that the computational complexity of a UE supporting MIMO directly translates into the UE's energy consumption. Because of all this, the academy and industry are each time more interested in having a high-fidelity PHY layer in network simulators such as \texttt{ns-3}. 

The ns-3 \texttt{nr} module (a.k.a., 5G-LENA) for simulation of 5G NR in sub 6 and \gls{mmwave} bands~\cite{PATRICIELLO2019101933} supports two types of beamforming methods: ideal and realistic, for analog beamforming. Ideal beamforming methods determine the beamforming vectors based on either the assumption of the perfect knowledge of the channel (e.g., cell scan method), or the exact positions of the devices (e.g., direction-of-arrival method), and they do not consume any time/frequency overhead to design the beamforming
vectors~\cite{mezzavilla2017end}. On the other hand, realistic beamforming methods, are expected to select the best beamforming based on some real measurements,
e.g., estimate the channel based on Sounding Reference Signals (SRSs)~\cite{bil21ns3}. To support spatial multiplexing and SU-MIMO, recently a Dual-Polarized MIMO (DP-MIMO) model was developed and implemented~\cite{bildpmimo22}. 
Such implementation is limited to a particular case of SU-MIMO, supporting only 2 streams per user with no possibility for extension. Indeed, it relies on some assumptions related to the mapping of the streams to ports that are limiting its application, i.e., each stream is mapped to the antenna elements with a specific polarization. This means that there is a hard limitation on the assignment of streams to a specific transmit port and the specific antenna elements of one polarization. In addition, it includes a not realistic inter-stream interference rejection model. 

Apart from 5G-LENA, MIMO has been considered in other ns-3 modules. The ns-3 \texttt{mmwave} module for simulation of 5G NR in mmWave bands~\cite{mezzavilla2017end}, supports beamforming and it has been recently extended to support MU-MIMO~\cite{9515585}. However, the \texttt{mmwave} module has been lacking up to now SU-MIMO, which is specially relevant for sub 6 GHz bands. In the area of IEEE, various MIMO models have been proposed~\cite{10.1145/3460797.3460799, 10.1145/3337941.3337947, 10.1145/3389400.3389403}. 
In the ns-3 \texttt{lte} module (a.k.a., LENA) there is an abstraction model to simulate 2$\times$2 SU-MIMO, which models up to two streams per user. However, such a model does not account for wireless propagation differences across different streams, assumes no correlation between antennas, and has a simplified inter-stream interference model, which are non-realistic assumptions for an accurate PHY modeling of SU-MIMO.

In this paper, we propose, implement, and evaluate a full SU-MIMO model for system-level simulation of 5G NR MIMO systems in \texttt{ns-3}, by extending the current framework with the support of MIMO spatial multiplexing to send multiple data streams towards the same user (SU-MIMO), thus enabling the hybrid beamforming feature of 3GPP. The code is publicly available and merged into the master branch of ns-3 5G-LENA module~\cite{5gLena}. As the analog beamforming feature has been widely analyzed in ns-3 previously, in this paper we focus on the new MIMO digital baseband processing. We have implemented codebook Type-I precoding (for single panel case), assuming MMSE-IRC (Minimum Mean-Squared Error Interference Rejection Combining) receiver for inter-stream interference suppression, as adopted in 3GPP. As compared to the old DP-MIMO implementation~\cite{bildpmimo22}, the new full MIMO implementation supports general SU-MIMO, through the inclusion of 3GPP-compliant PMI and RI reporting. Such general MIMO includes as a particular case the old DP-MIMO, but with a more generic and realistic model for DP-MIMO in which the streams/port/antennas mapping is not predefined but defined through digital precoding and in which a more accurate model for the inter-stream interference calculation is considered. The current model supports up to 32 antenna ports and rank 4. However, the full MIMO model can be easily extended to support more streams/ranks/ports.

The rest of the paper is organized as follows. Section~\ref{sec:mimo} reviews MIMO background, with special emphasis on the digital baseband processing and the support for MIMO in 3GPP 5G NR. Section~\ref{sec:FullMimo} details the implementation of the full MIMO system in ns-3 5G-LENA. Then, Section~\ref{sec:eval} presents the simulation results to validate the new MIMO model in ns-3. Section~\ref{sec:tradeoff} discusses the MIMO model's fidelity vs computational complexity in ns-3. Finally, Section~\ref{sec:conc} includes the conclusions and future work.

\section{MIMO background}
\label{sec:mimo}
In real systems, devices capable of performing MIMO spatial multiplexing can use more than one stream to transmit, e.g., a gNB in the downlink can send multiple streams to those UEs that support MIMO spatial multiplexing and can decode multiple streams simultaneously. MIMO technology has been known in textbooks for several decades, and it is used in 4G-LTE and 5G-NR and has been in Wi-Fi products for more than 20 years. For optimal MIMO performance, the gNB should apply a precoding matrix (at the transmitter) that determines how the signal is aligned relative to the channel, and the UE (at the receiver) should apply a receive filter to suppress inter-stream interference and recover each stream. Usually, the precoding matrix that provides maximum Signal-to-Interference-plus-Noise Ratio (SINR) for the given channel matrix is selected~\cite{palomar:phd} and the receive filter is usually designed to reduce the mean-square error between the transmitted and decoded data symbols, for which the MMSE-IRC receiver is usually adopted in 3GPP, as it provides a good balance between performance (mean-square error reduction) and implementation complexity (linear receiver).


\subsection{MIMO digital baseband processing preliminaries}\label{mimo:pre}
Consider a MIMO downlink system composed of $K$ gNBs, each serving a single UE on a given
frequency resource (resource block (RB) in 5G NR) and time resource (time slot in 5G NR).
Assume the $j$-th gNB ($j=1,\dots,K$) is equipped with $M_j$ antenna ports, and the $i$-th user ($i=1,\dots,K$) has $N_i$ antenna ports. Let $\textbf{H}_{i,j} \in \mathbb{C}^{N_{i} \times M_{j}}$ denote the channel matrix between the transmit antenna ports at the $j$-th gNB and the receive antenna ports at the $i$-th user~\footnote{Note that $\textbf{H}_{i,j}$ is the equivalent channel between transmit and receive antenna ports and so, with this formulation, it encapsulates the wireless propagation channel between all the transmit and receive antenna elements as well as the transmit and receive analog beamforming.}. Let $\textbf{P}_{i} \in \mathbb{C}^{M_{i} \times m_i}$ denote the transmit precoding matrix that the $i$-th gNB uses to transmit $m_i$ (data) streams towards its intended $i$-th user. To achieve multiple data stream transmission, the following constraints must be satisfied: $m_i\le M_i$ and $m_i\le N_i$.

Assuming narrow-band transmissions for a given frequency resource, the equivalent baseband
signal $\textbf{y}_i \in \mathbb{C}^{N_{i} \times 1}$ observed at the $i$-th user is:
\begin{equation}
\textbf{y}_i=\textbf{H}_{i,i}\textbf{P}_{i}\textbf{b}_{i}+\textbf{w}_{i}, \label{recSignal}
\end{equation}
where $\textbf{b}_{i}\in \mathbb{C}^{m_{i} \times 1}$ contains the symbols transmitted by the $i$-th gNB towards its serving ($i$-th) user, and $\textbf{w}_{i}$ is the noise-plus-interference received at the $i$-th user, i.e.,
\begin{equation}
\textbf{w}_i=\sum_{j=1,j \ne i}^K\textbf{H}_{i,j}\textbf{P}_{j}\textbf{b}_{j}+\textbf{n}_{i},
\end{equation}
where $\textbf{n}_{i}$ refers to the additive noise vector with covariance matrix $\textbf{N}_{i}$.
Assuming that interference is treated as noise and that a linear receive filter is applied at the user, the symbols are estimated at the $i$-th user as:
\begin{equation}
\hat{\textbf{b}}_i=\textbf{R}_{i}^H\textbf{y}_i,
\end{equation}
where $\textbf{R}_i \in \mathbb{C}^{N_{i} \times m_i}$ is the linear receive filter.

The mean square error (MSE) for the symbols transmitted towards the $i$-th user can be expressed through the so-called MSE-matrix $\textbf{E}_{i} = \mathbb{E}{\{(\hat{\textbf{b}}_i-\textbf{b}_i)(\hat{\textbf{b}}_i-\textbf{b}_i)^H\}} \in \mathbb{C}^{m_{i} \times m_{i}}$, being $\mathbb{E}\{.\}$ the expectation operator. The receive filter that minimizes the MSE, i.e., Tr($\textbf{E}_{i}$), being Tr(.) the trace operator, is the MMSE-IRC receiver~\cite{palomar:phd}. The MMSE-IRC receiver is given by: $\textbf{R}_{i}^\texttt{mmse}=(\textbf{H}_{i,i}\textbf{P}_{i}\textbf{P}_{i}^H\textbf{H}_{i,i}^H+\textbf{W}_{i})^{-1}\textbf{H}_{i,i}\textbf{P}_{i}$, where $\textbf{W}_{i}$ is the covariance of the noise-plus-interference received at the $i$-th user:
\begin{equation}
\textbf{W}_{i}=\sum_{j=1,j \ne i}^K\textbf{H}_{i,j}\textbf{P}_{j}\textbf{P}_{j}^H\textbf{H}_{i,j}^H+\textbf{N}_{i}. \label{W}
\end{equation}
Note that $\textbf{W}_{i}$ admits a Cholesky ``LLT" decomposition as follows: $\textbf{W}_{i}=\textbf{L}_{i}\textbf{L}_{i}^H$. This allows
transforming the received signal in Eq. \eqref{recSignal} through the following equivalent representation: $\textbf{y}_i^{\text{eq}}=\textbf{L}_{i}^{-1}\textbf{y}_i=\textbf{L}_{i}^{-1}\textbf{H}_{i,i}\textbf{P}_{i}\textbf{b}_{i}+\textbf{L}_{i}^{-1}\textbf{w}_{i}$, for which the second term ($\textbf{L}_{i}^{-1}\textbf{w}_{i}$) has an identity covariance matrix (also known as interference whitening process)~\cite{palomar:phd}. 

Using the MMSE-IRC receiver $\textbf{R}_{i}^\texttt{mmse}$, the MSE-matrix results~\cite{palomar:phd}:
\begin{equation}
\textbf{E}_{i} =(\textbf{I}+\textbf{P}_{i}^H\textbf{H}_{i,i}^H\textbf{W}_{i}^{-1}\textbf{H}_{i,i}\textbf{P}_{i})^{-1}=(\textbf{I}+\textbf{P}_{i}^H(\textbf{H}_{i,i}^{\text{intfNorm}})^H\textbf{H}_{i,i}^{\text{intfNorm}}\textbf{P}_{i})^{-1}, \label{E}
\end{equation}
where $\textbf{H}_{i,i}^{\text{intfNorm}}=\textbf{L}_{i}^{-1}\textbf{H}_{i,i}$ represents the interference-normalized channel. 

This way, 
the SINR of the $l$-th stream ($l=1,\dots,m_i$) of the $i$-th user is expressed as~\cite{palomar:phd}:
\begin{equation}
SINR_{i,l} =1/\textbf{E}_{i}(l,l)-1, \label{sinr}
\end{equation}
where $\textbf{E}_{i}(l,l)$ denotes the $l$-th diagonal element of the MSE matrix $\textbf{E}_{i}$ in Eq.~\eqref{E}~\footnote{The MSE computation Eq.~\eqref{E} has a bias~\cite{cioffi}, however, often channel decoder requires an unbiased estimate as input and for unbiased estimates the SINR values per stream is 1 less than the that of the biased estimate~\cite{cioffi}.}. Let us note that thanks to the introduction of $\textbf{H}_{i,i}^{\text{intfNorm}}$, we only need $\textbf{H}_{i,i}^{\text{intfNorm}}$ and $\textbf{P}_{i}$ to represent the signal for the $i$-th user and extract the SINRs for its streams.

\subsection{MIMO in 3GPP 5G NR}
In 3GPP, closed-loop MIMO is enabled thanks to the CSI feedback, which includes the PMI, the RI, and the CQI~\cite{TS38214}. The selection of the precoding matrix (i.e., $\textbf{P}_{i}$) that gives the best performance (i.e., maximum SINR) is done by the UE, and reported in the PMI through an index of a set of predefined precoding matrices from a codebook, as part of the CSI feedback message to the gNB. In addition, the UE also reports the RI (i.e., $m_i$) as part of the CSI feedback, which indicates to the gNB how many streams to use for that UE (i.e., $m_i$). Even if RI=1, we can still use precoding to combine the signals to/from multiple antenna ports in an optimal way. Each antenna port is then indeed aligned to the channel through analog beamforming. 3GPP 5G NR, different from LTE, allows sending up to 4 streams in the same transport block (TB). 

In 3GPP, the MIMO operations are enabled by the adoption of antenna arrays (usually modeled as dual-polarized linear antenna arrays~\cite{TR38901}) and the introduction of antenna ports concept, in which, for an antenna array of multiple antenna elements, multiple antenna elements are combined into one antenna port for digital processing (precoding), while analog processing (analog beamforming) is applied for the antenna elements within one antenna port. 3GPP 5G NR supports up to 32 ports, limited by the CSI-RS (CSI reference signals) design, over which the channel measurements are carried out. Antenna arrays are defined by a number of horizontal antenna elements, a number of vertical antenna elements in the physical antenna array, and whether the elements are dual-polarized or not, for the single panel case. Similarly, the number of virtual antenna ports in the horizontal and vertical directions ($N_h$ and $N_v$) are defined. In the case of cross-polarized antennas, the total number of ports for a device (gNB or UE) is given by: $2\times N_h \times N_v$. Table \ref{table:PortsConfig} shows the supported configurations for the number of ports in horizontal and vertical dimensions of the logical antenna array ($N_h$,$N_v$) specified by 3GPP, for the case of a single panel.

\begin{table}[!t]
\footnotesize
\centering
\caption{Supported antenna ports configurations in 3GPP}
\begin{tabular}{|p{3cm}|p{4cm}|}
 \hline
 Number of ports & ($N_h$,$N_v$) configurations \\
 \hline \hline
 2 & (1,1) \\
 4 & (2,1) \\
 8 & (2,2), (4,1)\\
 12 & (3,2), (6,1)\\
 16 & (4,2), (8,1) \\
 24 & (4,3), (6,2), (12,1) \\
 32 & (4,4), (8,2), (16,1)\\
 \hline
 \end{tabular}
\label{table:PortsConfig}
\end{table}

5G NR supports two types of codebooks: Type-I (with single-panel and multi-panel support) and Type-II. In the case of Type-I single-panel (which is the focus of this paper), the precoding matrix can be expressed as: $\textbf{P}_{i}=\textbf{P}_{i,1}\textbf{P}_{i,2}$, where $\textbf{P}_{i,1}$ includes wideband and long-term precoding and $\textbf{P}_{i,2}$ includes the subband and frequency-dependent precoding. $\textbf{P}_{i,1}$ is defined by a beam or group of beams pointing in various directions, while  $\textbf{P}_{i,2}$ chooses the group of vectors from $\textbf{P}_{i,1}$ and applies phase shifts over the polarizations. 


\section{New full SU-MIMO model for ns-3 and 5G-LENA} 
\label{sec:FullMimo}
The new full SU-MIMO model that we propose in this paper and that has been adopted recently in ns-3 and 5G-LENA can combine spatial multiplexing (with up to four streams per user, and 32 antenna ports) and beamforming (which applies for each of the streams). Multiple (up to four) streams are encoded in the same TB. PMI, RI, and CQI are implemented and included as part of the CSI feedback. It follows the 3GPP codebook-based Type-I model for precoding~\cite{TS38214} and assumes an MMSE-IRC receiver. For precoding and rank (PMI and RI) selection, an exhaustive search is implemented. The number of streams is called the rank in the code, which affects the TB size and other performance characteristics. The inter-stream interference is correctly computed through matrix processing, and this is why the use of more than 2 streams requires the Eigen library to compute operations like matrix inverse, singular value decomposition (SVD), etc. As the SINR and interference computations are correctly modeled, following~\cite{palomar:phd} and as described in Section~\ref{sec:mimo}, and multiple streams are fit in one TB, this allows using the SISO error model for MIMO error modeling, by vectorizing the 2D SINR (of dimensions number of RBs, rank) into 1D SINR (of dimensions number of RBs $\times$ rank, 1).

In the following, we explain the design choices and implementation details to enable the new full SU-MIMO in ns-3 and 5G-LENA. These include: 1) the extensions to ns-3 core, spectrum, and antenna modules to support 
the design and implementation of MIMO and the efficient operations with matrices (i.e., \texttt{MatrixArray}) (described in Sections~\ref{sec:max}-\ref{sec:ant}), and 2) the extensions performed in 5G-LENA (detailed in Sections~\ref{sec:ofdma}-\ref{sec:mimoAct}). Extensions in \texttt{ns-3} are two-fold; there were extensions to 
support the definition of multiple antenna ports and the spectrum channel matrix, and also the important set of changes was to improve the computational complexity of the operations that are needed on complex 3D data structures that represent channel matrices in time and frequency. The changes in the \texttt{nr} module include 1) the removal of ``the OFDMA downlink trick",  2) the MIMO interference and SINR calculations, as well as the interfaces to pass the results to other classes, 3) the computation of transport block error rate (TBLER) based on the MIMO SINR, 4) implementation of the 3GPP-compliant Type-I precoding codebooks, 5) the optimal precoding search algorithm, which the UE needs to send as a feedback to the gNB in the PMI, 6) adding the ``rank" parameter to many interfaces throughout the code, and 7) the enabling of the new methods for configuration and enabling the MIMO (i.e., enabling the MIMO feedback). All these extensions to the 5G-LENA module have been included in the 5G-LENA mainline in its Release 3.0, except for the 32-port codebook, which is to be released shortly in the 5G-LENA 3.1 Release \cite{5gLena}.

\subsection{Introducing MatrixArray into ns-3}\label{sec:max}
The 3GPP channel matrix in ns-3 implements a 3D structure whose dimensions depend on the number of the transmit antenna, the receive antenna, and the number of clusters. Such a structure models the channel in a time domain, i.e., there is a channel matrix per cluster. To extend the 3GPP channel model to support full MIMO, we need to consider the frequency domain channel matrix, having three dimensions: the number of receive ports, the number of transmit ports, and the number of RBs, i.e., as mentioned in Section \ref{mimo:pre}, $\textbf{H}_{i,j} \in \mathbb{C}^{N_{i} \times M_{j}}$, i.e., there is a $\textbf{H}_{i,j}$ per RB. Hence, when moving to MIMO design there is a shift of the heavy computations from time to frequency domain. While in the previous ns-3 3GPP channel model implementation the third dimension is relatively small, e.g., the number of clusters is between 10 and 20, the third dimension of the frequency domain channel matrix can be much larger, e.g. 275 RBs. During the development of MIMO models in ns-3, we found out that this third dimension can affect significantly the computational performance when using the 3D data structures that were available in ns-3 before our contribution, i.e., to represent such 3D structures ns-3 and its spectrum module used a C++ std vector of a vector of a vector. Pagin at al. in~\cite{pagin} has improved the efficiency of the 3GPP channel model in ns-3 by allowing the usage of Eigen~\cite{eigen} to perform computations on 3D structures. Thanks to that upgrade when Eigen is available the 3GPP channel is represented as \texttt{std::vector} of Eigen matrices. This improved the performance of the 3GPP channel model significantly. However, we found out that this improvement was not enough for MIMO computations. We observed that using such vector of matrices (regardless they are C++ or Eigen) becomes inefficient when the third dimension increases which is the case with the frequency domain matrices needed for MIMO models. For this reason, we have investigated and proposed to use a new data type in ns-3 that can provide more efficient storage, access, and linear algebra operations for 3D structures, i.e., arrays of matrices, which are typically used either in the current SISO or future MIMO models. In this respect, we proposed a new class called ValArray that leverages std::valarray to represent 3D structures. We extended this interface, by an additional class called \texttt{MatrixArray}, which treats this 3D array, as an array of matrices, on which linear algebra operations can be performed page by page (matrix by matrix). We created tests for \texttt{ValArray} and \texttt{MatrixArray} to test all the implemented functionalities. Apart from proposing a more efficient representation of 3D structures for SISO and MIMO, \texttt{MatrixArray} has improved significantly the interfaces that are using Eigen because a single class is a unique wrapper for both implementations, the one using Eigen, and one that is not. This is a much better approach than having two independent definitions of the 3D structures, as it was implemented in ns-3 before our changes. 
The proposed \texttt{MatrixArray} approach has passed the ns-3 review process and it was included in the official ns-3 since ns-3.38 Release \cite{ns-3}.

\begin{figure}[!t]
\centering
\subfloat[ns-3 3D arrays without Eigen]{\includegraphics[width=0.49\textwidth]{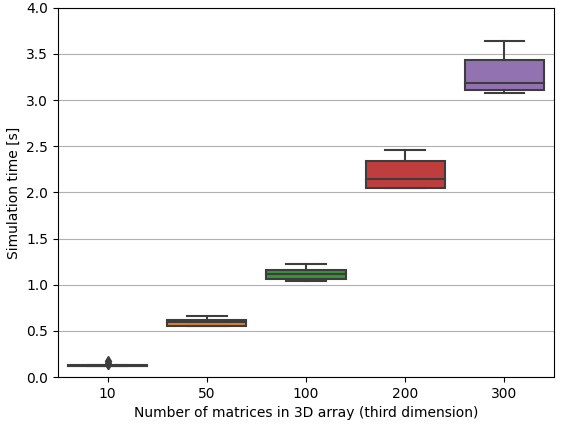}}
\subfloat[ns-3 3D arrays with Eigen]{\includegraphics[width=0.49\textwidth]{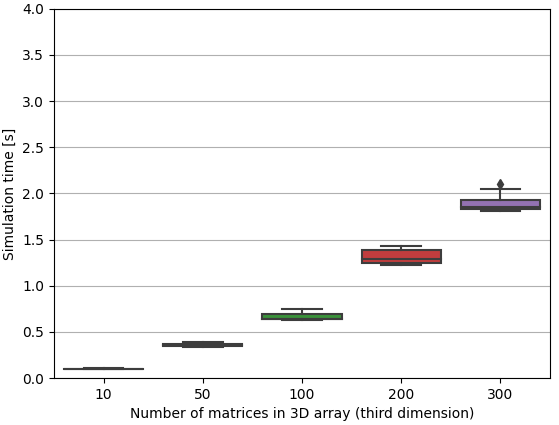}}
\\
\subfloat[MatrixArray without Eigen]{\includegraphics[width=0.49\textwidth]{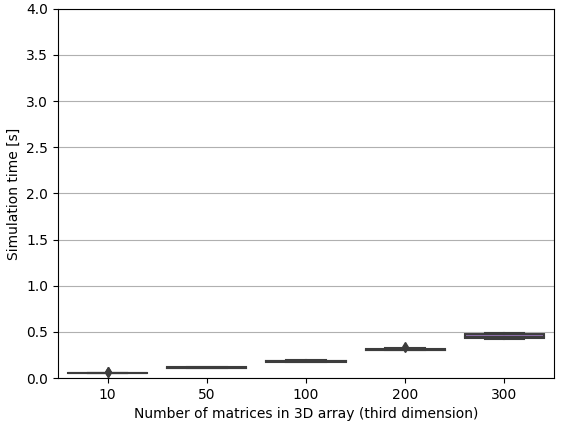}}
\subfloat[MatrixArray with Eigen]{\includegraphics[width=0.49\textwidth]{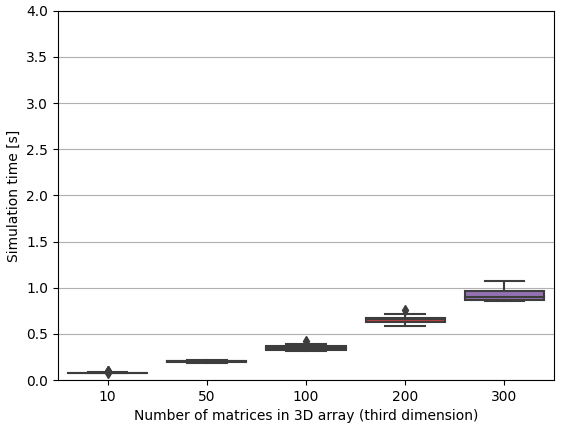}}
\\
\caption{Comparison of the computational complexity of the ns-3 3D array data structure based on the std::vector of matrices and the proposed MatrixArray-based approach} 
\label{matrix-array-perf}
\end{figure}

We performed profiling to demonstrate that the proposed approach is notably more efficient than other approaches when the third dimension increases. To perform profiling we created an example that performs 1e5 multiplications of two 3D arrays and assigns the value to the third 3D array. We test different values for the third dimension (10, 50, 100, 200, 300). Matrices in the 3D array are of 2x2 dimensions. Profiling is carried out by using the optimized mode. In Figure~\ref{matrix-array-perf} are shown the benchmark results. We observe that depending on the setup the proposed \texttt{MatrixArray}-based approach is performing equally well or better. The major improvement comes from the usage of underlying \texttt{std::valarray} to represent the 3D structure instead of the usage of std::vector of matrices. The usage of C++ \texttt{std::valarray} instead of the C++ \texttt{std::vector} of the vector of vectors allows for lower overheads when accessing the values, i.e., single direct access versus the access operation composed of 3 redirections. Also, the usage of a single \texttt{std::valarray} guarantees that all the data belonging to the same matrix is in the contiguous memory allocations, which significantly reduces the overheads of loading and caching, i.e., the efficiency of each load of the fragment of memory increases concerning the C++ std vector-based 3D structures, where we load parts of memory that are not necessary for the current matrix operation.
Notice that \texttt{MatrixArray}-based approach when using Eigen leverages the \texttt{Eigen::Map} function which uses directly the piece of memory where the specific matrix is placed, which allows us to use the unique \texttt{MatrixArray} interface for both cases,  with and without Eigen. We do, however, notice that there is an overhead of using Eigen when the matrices are of small dimensions, such as 2x2, as in this profiling example.

\subsection{Extending ns-3 antenna and spectrum modules to support MIMO}\label{sec:ant}

In the ns-3 \texttt{antenna} module is added support for dual-polarized antenna arrays and for multiple ports in the antennas, which are 
a prerequisite to support a full MIMO model implementation. 
\texttt{UniformPlannarArray} class that models the uniform planner arrays is extended to support the concept of antenna ports following the sub-array partition model for TXRU virtualization, as described in Section 5.2.2 of 3GPP TR 36.897~\cite{TR36897}. In case the antenna array is dual polarized, the total number of antenna elements is doubled and the two polarizations are overlapped in space.

On the other hand, the 3GPP channel model in the ns-3 \texttt{spectrum} module is extended to support multiple ports, dual-polarized antenna elements, and the generation of the frequency domain spectrum channel matrix. In MIMO systems multiple transmit and receive antenna ports can exist, hence the power spectral density (PSD) is multidimensional. It was not anymore enough to represent a signal as before with a single PSD, instead, the frequency domain channel matrix is introduced to represent the channel. The dimensions of such a frequency domain channel matrix are the number of receive antenna ports, the number of transmit antenna ports, and the number of subbands in frequency (or RBs). Additionally, the 3D precoding matrix is introduced to be able to correctly perform the MIMO calculations of the received signal and interference. The dimensions of the precoding matrix are the number of transmit antenna ports, the number of transmit streams, and the number of subbands (or RBs).
To achieve this, \texttt{CalcRxPowerSpectralDensity} is not returning a single \texttt{Ptr<SpectrumValue>} as before but is returning \texttt{Ptr<SpectrumSignalParameters>} which is extended to include the frequency domain spectrum channel matrix and the precoding matrix. Finally, ns-3 \texttt{ThreeGppChannelModel} is extended to support the calculation of the MIMO channel matrix by taking into account multiple ports and/or dual-polarization.
When multiple ports are being configured the sub-array partition model is adopted for TXRU virtualization~\cite{TR36897}, and so equal beam weights are used for all the ports. Support of the full-connection model for TXRU virtualization would need extensions.
To preserve backward compatibility, SISO 3GPP channel calculations are still performed as before these changes for MIMO.
To achieve this, i.e., to compute a single received PSD, the frequency domain 3D spectrum channel matrix is used. In the case of multiple ports at the transmitter, the received PSD is calculated
by summing per each RB the real parts of the diagonal elements of the $(\textbf{H}_{i,i}\textbf{P}_{i})^H * (\textbf{H}_{i,i}\textbf{P}_{i})$, where $\textbf{H}_{i,i}$ is the frequency domain spectrum channel matrix and $\textbf{P}_{i}$ is the precoding matrix (see Section~\ref{mimo:pre}).
To test these new MIMO extensions in the antenna and spectrum module, we developed two new additional tests: 
\begin{itemize}
\item \texttt{ThreeGppCalcLongTermMultiPortTest} which tests that the channel matrices are
  correctly generated when multiple transmit and receive antenna ports are used, and 
\item \texttt{ThreeGppMimoPolarizationTest} which tests that the channel matrices are
  correctly generated when dual-polarized antennas are being used.
\end{itemize}
These extensions to support full MIMO have been included in the ns-3 mainline in the ns-3.41 Release~\cite{ns-3}.

\subsection{Removal of the ``OFDMA downlink trick"}\label{sec:ofdma}
The module was initially designed with TDMA as the primary scheduling mechanism for downlink transmissions from a single gNB (base station) to multiple UEs (user equipment, e.g., a mobile phone) within a single time slot. However, when OFDMA was introduced, the model employed a technique called ``OFDMA downlink trick" to combine multiple transmissions into a single one. This approach posed challenges for a seamless MIMO design in the nr module. Consequently, eliminating the ``OFDMA downlink trick" became the main prerequisite for a clean MIMO design and implementation.      

In TDMA, the 5G-LENA model is straightforward. Packets destined for a specific UE are combined into a single packet burst, creating an independent representation for each signal to model the transmission. This maintains a one-to-one relationship between the data packets sent to each UE (packet burst), the downlink allocation specified in Downlink Control Information (DCI), which contains metadata about the transmission, and the physical layer signal transmitted from the antenna. This relationship is ensured by considering the starting symbol (symStart) as an additional parameter when mapping packets to packet bursts.

However, in OFDMA, the ``OFDMA downlink trick" disrupts this one-to-one relationship. Now, there is only one packet burst containing all packets for all UEs, as symStart is the same for all UEs. Because of this, only the downlink allocation with metadata for the first transmission (e.g., DCI for UE 1) is used, with additional logic implemented to skip the other DCIs. Also, the physical layer signal is created as a combination of all allocations. OFDMA downlink trick poses challenges for MIMO design, as gNB may transmit towards various UEs using different ranks. In such cases, the OFMDA downlink trick requires merging the representations of the different MIMO signals, which is difficult to implement and error-prone. For this reason, the OFDMA downlink trick technique is removed.
To implement this removal, RNTI is introduced as an additional parameter in the packet burst mapping. This modification establishes a one-to-one correspondence between the downlink allocation or DCI towards each UE and the respective packet burst.

We removed the logic that was skipping all the downlink allocation transmissions other than the first, i.e., after the removal of the trick, all downlink locations are transmitted in independent signal representations.
To avoid incorrect modeling of analog beamforming, we added logic to ensure that the analog beamforming vector is changed only once per transmission event.
We removed the combining of all allocations into a single physical signal. We kept parts of the existing combining algorithm to ensure the correct scaling of the transmit PSD.

Notice that the removal of the ``OFDMA downlink trick" is a step forward towards the implementation of the full OFDMA in 5G-LENA without having the current limitation of scheduling at the same time only the UEs belonging to the same beam. The current implementation schedules UEs with the same analog beam, which applies to all ports (and so all streams) of each UE, but different digital beams can be used through digital processing and the precoding matrix $\textbf{P}_i$, which is designed per UE and can vary across RBs. Since now, after the removal of this trick, the signal to each UE is independently transmitted, for each of them, and gNB can use a different digital beam on each RB. For this full OFDMA to work correctly, the OFDMA schedulers in 5G-LENA should be updated to not limit the scheduling at the same time of the UE's belonging to the same analog beam.

\subsection{The MIMO interference and SINR calculations}
The \texttt{nr} model for the interference calculation is extended to support the calculation of the MIMO interference and
MIMO SINR calculations. The main class for the calculation of the interference in the NR module is \texttt{NrInterference} class.
This class is extended with new functions for the computation of the interference-and-noise covariance matrix and SINR.
These functions are \texttt{CalcOutOfCellInterfCov}, \texttt{CalcCurrInterfCov}, \texttt{AddInterference}, and \texttt{ComputeSinr}.
\texttt{CalcOutOfCellInterfCov} computes the interference signals from all out-of-cell interferers.
\texttt{CalcCurrInterfCov} prepares \texttt{NrInterference} class for MU-MIMO by supporting the calculation of the interference signals
by considering the interferers from the same cell. For example, in the uplink MU-MIMO, UEs from the same cell could act as interferers.
\texttt{AddInterference} adds the covariance of the signal to an existing covariance matrix.
Finally, \texttt{ComputeSinr} computes the SINR as follows:
\begin{enumerate}
\item the interference-plus-noise covariance matrix of the received signal is computed (i.e., $\textbf{W}_i$ in Eq. \eqref{W}),
\item the received signal is transformed into an equivalent representation where the interference-plus-noise covariance is an identity matrix (a.k.a., interference whitening transformation, see Section~\ref{mimo:pre}) and $\textbf{H}_{i,i}^{\text{intfNorm}}=\textbf{L}_{i}^{-1}\textbf{H}_{i,i}$ is computed, 
\item a dummy precoding matrix $\textbf{P}_i$ is created when none exists, and
\item the SINR of each stream based on the MSE matrix is computed as detailed in Eq. \eqref{E}, by using $\textbf{H}_{i,i}^{\text{intfNorm}}$ and $\textbf{P}_i$.
\end{enumerate}

To support all these MIMO operations, it was not enough to use a single dimensional \texttt{SpectrumValue} type that has been traditionally used in \texttt{NrInterference} for SISO. To support efficient storage and computations of MIMO operations new classes were defined, such as \texttt{NrCovMat}, \texttt{NrIntfNormChanMat} and \texttt{NrSinrMatrix}. \texttt{NrCovMat} stores the interference-plus-noise covariance matrices of a MIMO signal, with one matrix page for each frequency bin. This class also provides some functions for efficient computations on covariance matrices. Its function \texttt{CalcIntfNormChannel} performs interference whitening (see Section~\ref{mimo:pre}).
\texttt{NrIntfNormChanMat} stores the interference-whitened channel matrix, the channel matrix after normalizing/whitening the interference, i.e., $\textbf{H}_{i,i}^{\text{intfNorm}}=\textbf{L}_{i}^{-1}\textbf{H}_{i,i}$. Its function \texttt{ComputeSinrForPrecoding} computes the SINR based on MSE, see Eq.~\eqref{sinr}. Finally, \texttt{NrSinrMatrix} stores the MIMO SINR matrix whose dimensions are the rank and the number of RBs. 

The MIMO interference and SINR calculations require Eigen3~\cite{eigen}, a C++ template library for linear algebra: matrices, vectors, numerical solvers, and related algorithms. However, the Eigen library is not always available. To allow the compilation even when Eigen is not available a CMake switch is added:

  a) when Eigen is enabled, the file nr-mimo-matrices-eigen.cc is compiled
  
  b) when Eigen is disabled, the file nr-mimo-matrices-no-eigen.cc is compiled (the implementations just contain a single NS\_FATAL\_ERROR). In this case, users can still compile but can only use SISO. The functions used from the Eigen library could be in the future implemented in ns-3 to reduce the dependency of the nr module on the Eigen library. Then nr-mimo-matrices-no-eigen.cc could be implemented to call these ns-3 alternatives of Eigen functions.

To support the multi-dimensional MIMO signals a new interference chunk processor called \texttt{NrMimoChunkProcessor} is introduced. This class mirrors the original \texttt{LteChunkProcessor} that is originally used in \texttt{NrInterference} for SISO.
\texttt{LteChunkProcessor} is not sufficient for MIMO because it can only store a frequency-domain vector of SINR values whereas MIMO requires a 2D matrix with the dimensions: number of RBs and number of MIMO streams. \texttt{LteChunkProcessor} stores the sum of the different signals' power spectral density values and performs the averaging once the function \texttt{End} is called. Such SINR averaging in the time domain limits the fidelity. In general, each received signal may have a different number of MIMO layers, hence combining the SINR of different signals is not trivial. To avoid all this, \texttt{NrMimoChunkProcessor} keeps a list with full information of all different signals, and no averaging is performed at this stage. The averaging now must be implemented later in the error model which opens the door also for different possible implementations, e.g., the error model may apply exponential effective SINR both over time and frequency. Hence \texttt{NrMimoChunkProcessor} only looks like \texttt{LteChunkProcessor} but is mainly used as a storage to pass
the information to other entities that can perform different computations by exploiting the full information of all different signals. \texttt{NrMimoChunkProcessor} provides two kinds of callbacks:
\begin{itemize}
\item Providing MIMO SINR: one 2D matrix for each different time-domain chunk, which is used by the error model to compute TBLER, and   
\item Providing the interference covariance matrices for each different time-domain chunk are passed to CQI feedback function and are used together with the channel matrix to compute the precoding matrix PMI feedback.
\end{itemize}

\subsection{Computation of TBLER based on the MIMO SINR}
A new function called \texttt{GetTbDecodificationStatsMimo} is added to \texttt{NrErrorModel} to determine if a transport block was received successfully. \texttt{GetTbDecodificationStatsMimo} performs a simple weighted average over potentially multiple different signal values received over time to get a single SINR matrix. The SINR matrix is then linearized to a vector and passed to the existing error model for SISO by calling a function \texttt{GetTbDecodificationStats}. Effectively, \texttt{GetTbDecodificationStatsMimo} is a
translation layer between the new MIMO code and the existing SISO error model.

When using MIMO one should configure the \texttt{AmcModel} as \texttt{ErrorModel}. The \texttt{ShannonModel} is not yet supported with MIMO, some additions are needed to \texttt{NrAmc} to allow its usage.

\subsection{Implementation of the 3GPP-compliant Type-I precoding codebooks}
The main base class for the implementations of Type-I codebooks is called \texttt{NrCbTypeOne}. This is a wrapper class for the implementations of Type-I precoding matrices in 3GPP TS 38.214. The main functions that \texttt{NrCbTypeOne} offers are: \texttt{Init}, \texttt{GetNumI1}, \texttt{GetNumI2} and \texttt{GetBasePrecMat(i1,i2)}. \texttt{Init} is used to initialize the codebook parameters based on the configuration through the attribute values. \texttt{GetI1} returns the number of wideband precoding indices i1. Similarly, \texttt{GetI2} returns the number of subband precoding indices i2. Finally, \texttt{GetBasePrecMat(i1,i2)} returns the 2D precoding matrix of size gNB ports x rank. While \texttt{NrCbTypeOne} represents only the interface, we provide two 3GPP-compliant specializations of this class: \texttt{NrCbTwoPort} and \texttt{NrCbTypeOneSp}.

\texttt{NrCbTwoPort} provides the implementation of the two-port codebook as per 3GPP TS 38.214. This class implements a codebook for a gNB with at most 2 antenna ports. For a single port, it returns a single-element matrix with a value of 1.0. For two ports, it implements Table 5.2.2.2.1-1~\cite{TS38214}: Codebooks for 1-layer and 2-layer CSI reporting using antenna ports 3000 to 3001. 

\texttt{NrCbTypeOneSp} provides the implementation of Type-I Single-Panel Codebook 3GPP TS 38.214 (Rel. 15, Sec. 5.2.2.2.1~\cite{TS38214}) that supports up to 32 antenna ports. It supports codebook mode 1 only and the rank is limited to 4. Codebook mode 1 means the per-subband i2 beam index is used only for the phase shift of the second polarization, while codebook mode 2 would use i2 also for beam refinement. \texttt{NrCbTypeOneSp} extends the \texttt{NrCbTypeOne} interface significantly by adding many new functions for the initialization of different parameters, such as i11, i12, i13, k1-k2 for different ranks, and the initialization of the full precoding matrix $\textbf{P}_{i}$. 
\texttt{CreateUniqueBfvs} is implemented to 
create a list of different (digital) beamforming vectors used for the first polarization. When the number of ports is 2, this function returns only a single element of value 1.0. Otherwise, it returns:

\begin{itemize}    
\item for rank 1, the vector $v_{l,m}$ (TS 38.214 Table 5.2.2.2.1-5~\cite{TS38214}), 
\item for rank 2, the two vectors $v_{l,m}$ and $v_{l',m'}$ (TS 38.214 Table 5.2.2.2.1-6~\cite{TS38214}), 
\item for rank 3 or 4 with less than 16 antenna ports, the two vectors $v_{l,m}$ and $v_{l',m'}$ (these are the unique vectors in the upper parts of TS 38.214 Tables 5.2.2.2.1-7 and 5.2.2.2.1-8~\cite{TS38214}), and finally, 
\item for rank 3 or 4 with at least 16 antenna ports, the two concatenated vectors $[\tilde{v}_{l,m}; \theta_p \times \tilde{v}_{l,m}]$ and $[\tilde{v}_{l,m}; -\theta_p \times \tilde{v}_{l,m}]$.
\end{itemize}
Additional functions are added to support the codebook implementation, such as, among others, \texttt{Kroneckerproduct} to create a Kronecker product of two vectors, e.g., used when creating the vector $v_{l,m}$ and ${\Tilde{v}_{l,m}}$. Also, \texttt{GetNumI11}, \texttt{GetNumI12}, and \texttt{GetNumI13} are added to provide the number of i11, i12, and i13 indices which represent, respectively, the horizontal and the vertical beam directions, and co-phasing shifts for a secondary beam. Notice that $i1$ is defined as a vector in~\cite{TS38214}. To reduce the number of loops and parameters, this vector is mapped to a unique integer. The mapping that we propose is as follows: i1 is created from i11, i12, i13 as $i1 = i11 + numI11 * (i12 + numI12 * i13)$, where $numI11$ is the number of i11 values (horizontal beam indices) and $numI12$ is the number of i12 values (vertical beam indices). This mapping is an arbitrary choice, and the ordering of this mapping does not impact a full search over all indices. The functions, \texttt{MapToI11(i1)}, \texttt{MapToI12(i1)}, and \texttt{MapToI13(i1)} are created to map the composite i1 index to an i11 (horizontal beam direction), i12 (vertical beam direction) and i13 (co-phasing of a secondary beam) indexes, respectively. When the function \texttt{GetBasePrecMat(i1,i2)} is called, then the indices i11, i12 and i13 are obtained by using these mapping functions and \texttt{GetBasePrecMatFromIndex(i11,i12,i13,i2)} function is 
called that returns the precoding matrix for such indices.
Additional Type-I codebooks, e.g., codebook mode 2, could be implemented by defining a new specialization of \texttt{NrCbTypeOne} codebook.


\subsection{Search for the optimal precoding matrix}
\label{pm-search}
\texttt{NrPmSearchFull} class is implemented to find the optimal precoding matrix, rank indicator, and corresponding CQI, and creates a CQI/PMI/RI feedback message. \texttt{NrPmSearchFull} uses exhaustive search for 3GPP Type-I codebooks.
Optimal rank is considered as the rank that maximizes the achievable TB size when using the optimal PMI. To determine the rank indicator the algorithm loops through all ranks (the number of MIMO streams), and for each rank it computes PMI, and it computes the maximum supported MCS, and associated TB size, and finally, it selects the rank and PMI that result in the highest TB size.
The optimal wideband (WB) and subband (SB) PMI values are periodically updated based on the configured update intervals, which can be configured using two attributes in the \texttt{NrUePhy} class: \texttt{WbPmiUpdateInterval} and \texttt{SbPmiUpdateInterval}. 

\begin{algorithm}[!t]
\begin{algorithmic}
\caption{Exhaustive PM search} 
\label{alg1}
\small
\State input: Interference-normalized channel matrix 
\For{$<Each~rank \in R>$}
  \For{$<Each~i1 \in I1>$}
  \For{$<Each~sb \in S>$}
   \For{$<Each~i2 \in I2>$}
        \State $C_{sb}(sb,i1,i2,rank) \gets ComputeCapacityForPrecoders(H(sb)\times P(i1, i2, rank))$        
         \If{$C(sb, i1, i2, rank) > C_{sb}(sb, i1, bestI2(sb), rank)$}
            \State $bestI2(sb) \gets i2$
         \EndIf 
   \EndFor
   \State $C_{wb}(i1, rank) = C_{wb} + C_{sb}(sb, i1, bestI2(sb), rank)$
  \EndFor  
  \If{$C_{wb}(i1, rank) > C_{wb}(bestI1, rank)$}
      \State $bestI1(rank) \gets i1$ 
   \EndIf
 \EndFor
  \If{$C_{wb}(bestI1(rank), rank) > C_{wb}(bestI1(rank), bestRank)$}  
        \State $bestRank \gets rank$ 
   \EndIf
\EndFor
\State output: bestRank, bestI1, bestI2 (per each subband)
\end{algorithmic}
\end{algorithm}

When a PMI update is requested, the optimal precoding matrices are updated using an exhaustive search over all possible precoding matrices specified in a codebook compatible with 3GPP TS 38.214 Type-I. The exhaustive search loops over all possible subband precoding matrices computes the SINR that would be achieved by each precoding matrix
and selects the precoder resulting in the highest average SINR. The full search algorithm is illustrated in Algorithm \ref{alg1}. Finally, the feedback message is created that includes the optimal rank, the corresponding optimal PMI, and CQI.

\texttt{NrPmSearch} is the base class and \texttt{NrPmSearchFull} is one possible specialization that finds PMI, RI, and CQI by using the described exhaustive search. One could create another specialization of \texttt{NrPmSearch} that would implement a different algorithm to find PMI and RI values.

\subsection{Rank parameter}
Many interfaces are extended in 5G-LENA to allow passing the rank number (the number of MIMO streams).
In this way, for example, already existing functions for the calculation of the transport block size for SISO could be easily updated to be used for both, SISO and MIMO (in this section we refer to SISO as the single stream transmission, although multiple antennas are supported, and MIMO for the multiple-stream transmission). Also, packet traces are extended to include the rank number.

\subsection{MIMO activation}\label{sec:mimoAct}
\texttt{NrHelper} class is responsible for setting the \texttt{NrPmSearch} algorithm to \texttt{NrUePhy} instance, and the configuration of the corresponding parameters, such as the type of the search algorithm, the type of the codebook, and the rank limit. \texttt{NrHelper} also creates \texttt{NrMimoChunkProcessor}
and adds the necessary callbacks. These callbacks are:

\begin{itemize}
    \item \texttt{NrSpectrumPhy::UpdateMimoSinrPerceived} which is called to provide MIMO SINR feedback
    \item \texttt{NrUePhy::GenerateDlCqiReportMimo} which is called to provide MIMO signal to functions that perform PMI search and create CQI/PMI/RI feedback
\end{itemize}

To enable MIMO in the simulation, the \texttt{EnableMimoFeedback} attribute of the \texttt{NrHelper} should be set to true. To configure PMI search parameters (such as rank limit, PMI search method, and the codebook) \texttt{NrHelper} provides a function \texttt{SetupMimoPmi}.
The \texttt{EnableMimoFeedback} enables MIMO feedback including PMI/RI/CQI, while \texttt{RankLimit} limits the possible RI value (e.g., to 1 stream). So, even if RI is limited to 1, the usage of MIMO feedback can provide benefits because of the PMI feedback.
 
\section{Evaluation and analysis}
\label{sec:eval}
In this section, we evaluate and confirm the correct behavior of the newly developed full SU-MIMO model, through multiple simulation campaigns over various antenna array configurations.

\begin{table}[!t]
\scriptsize
\centering
\caption{Scenario configuration parameters}
\begin{tabular}{|p{5cm}|p{5cm}|}
\hline 
Parameter name& Parameter Value \\ \hline \hline
ns-3 simulator release & Release 41 \\ \hline
5G-LENA NR module & Release v3.0~\cite{5gLenaRelease3} \\ \hline
Independent random simulation runs & 100 repetitions per setup (ns-3 RNG values: 1-100)\\ \hline
Simulation time & 1 seconds \\ \hline
CBR traffic inter-packet interval time & $40{\mu}s$ \\ \hline
CBR traffic packet size & 1000 bytes \\ \hline
Channel model  & 3GPP SCM TR 38.901 UMa \\ \hline
Channel update period & 100~ms \\ \hline
LoS/NLoS condition model & 3GPP TR 38.901 UMa \\ \hline
Los/NLoS update period & 100~ms \\ \hline
Central frequency & 4 GHz \\ \hline
Bandwidth & 10 MHz \\ \hline
gNB transmit power & 41 dBm \\ \hline
gNB and UE antennas & UPA 3GPP~\cite{TR38901} \\ \hline
MIMO codebook & Two-port and 32-port codebook~\cite{TS38214} \\ \hline
WB and SB update intervals & 10 ms and 2ms \\ \hline
gNB and UE number of panels & Single-Panel \\ \hline
gNB and UE antenna spacing & $d_h = 0.5 \lambda$, $d_v = 0.5 \lambda$ \\ \hline
gNB and UE bearing angles & 0 and 180 degrees \\ \hline
gNB and UE polarization slant angles & 0 and 90 degrees \\ \hline
gNB and UE antenna height & 25 and 1.5 meters \\ \hline
gNB and UE noise figure & 5 dB and 7 dB\\ \hline
gNB and UE antenna gain & 8 dBi (max) and 0 dBi\\ \hline
Numerology & 0 (15 kHz subcarrier spacing) \\ \hline
Duplexing mode & TDD \\ \hline
Multiple access scheme & TDMA \\ \hline
TDD subframe pattern & Flexible ($\text{F}|\text{F}|\text{F}|\text{F}|\text{F}|\text{F}|\text{F}|\text{F}|\text{F}|\text{F}$) \\ \hline
TDD slot format & 19 ($\text{DL}|\text{F}|\text{F}|\text{F}|\text{F}|\text{F}|\text{F}|\text{F}|\text{F}|\text{F}|\text{F}|\text{F}|\text{F}|\text{UL}$) \\ \hline
MCS Table & Table 2, TS 38.214 Table 5.1.3.1-2 \\ \hline
HARQ combining method & Incremental redundancy \\ \hline
Max no. of HARQ processes & 20 \\ \hline
Max no. of HARQ retx & 3 \\ \hline
UE processing delay (K1) & 2 slots \\ \hline
RLC mode & Unacknowledged mode \\ \hline
\end{tabular}
\label{mimo_scenario_configuration}
\end{table}

\subsection{Evaluation scenario}
To evaluate the MIMO implementation we created a simple topology consisting of one gNB and one UE~\footnote{An additional pair of gNB and UE can be enabled to simulate the impact on the interference. Such scenario configuration will be used in Section~\ref{eval-interf}.}.
The UE receives the downlink constant bit rate (CBR) traffic from the remote host. The default scenario parameters, such as the central frequency, bandwidth, gNB transmit power and down-tilt angle, and gNB and UE antenna heights, are set up according to a typical 3GPP urban macro dense scenario, such as given in Table 10 in~\cite{R1-1707360}. In Table~\ref{mimo_scenario_configuration}, we list all the simulation parameters, except for the specific number of gNB and UE antenna array elements, antenna ports, and polarizations that depend on the specific MIMO configuration being evaluated, which we provide in Table~\ref{antenna_scenario_conf} and will specify in the following sub-sections which specific configuration is used in each simulation campaign. The antenna port configurations used in simulation campaigns are according to those supported by 3GPP, as shown in Table~\ref{table:PortsConfig}. We define 5 different simulation campaigns to evaluate different aspects of MIMO implementation. These simulation campaigns are created to analyze:
\begin{enumerate}
\item MIMO implementation using a single-panel 2-port codebook, in which we analyze the impact of using the precoding matrix versus a base-line SISO antenna and 2-stream MIMO in Section~\ref{res-single-panel-2-port}, 
\item MIMO implementation using a single-panel 32-port codebook, in which we analyze how different gNB antenna configurations result in different rank selections and achieved performance (Section~\ref{res-single-panel-32-port}), 
\item the impact of gNB antenna ports configuration when using single-panel 32-port codebook (Section~\ref{res-antenna-ports}), 
\item the impact of SB and WB PMI interval update on the achieved throughput and the simulation execution time when using a single-panel 32-port codebook (Section~\ref{res-sb-wb-intervals}), and 
\item MIMO implementation when an interfering gNB and UE are present in the scenario (Section~\ref{eval-interf}).
\end{enumerate}
In each simulation campaign, we vary the distance between gNB and UE, i.e., by moving the UE away from gNB in the x-axis direction. We collect the measurements of the end-to-end application layer throughput, delay, jitter, selected MCS, rank, and execution time. We are interested not only to understand the performance improvements but also to understand the impact of the MIMO implementation on the computational complexity of the simulations. Finally, for each specific scenario configuration (e.g., antenna configuration, distance between gNB and UE, etc.), we run 100 independent random repetitions to produce statistically confident results.

\begin{table}[!t]
\scriptsize
\caption{Antenna configuration parameters for different simulation campaigns}
\centering
\begin{tabular}{|c|c|c|c|c|c|c|}
\hline 
Antenna configuration & Device & $N_{h}$ & $N_{v}$ & Rows & Cols & X-polarized   \\ \hline \hline
Conf 1 & gNB & 2 & 1 & 2 & 4 & No \\ \hline
Conf 1 & UE & 2 & 1 & 2 & 2 & No \\ \hline
Conf 2a & gNB & 2 & 2 & 4 & 2 & Yes \\ \hline
Conf 2b & gNB & 4 & 2 & 8 & 4 & Yes \\ \hline
Conf 2 & UE & 2 & 1 & 1 & 2 & Yes \\ \hline
Conf 3 with 8 ports & gNB & 2 & 2 & 8 & 4 & Yes \\ \hline
Conf 3 with 16 ports & gNB & 4 & 2 & 8 & 4 & Yes \\ \hline
Conf 3 with 32 ports & gNB & 4 & 4 & 8 & 4 & Yes \\ \hline
\end{tabular}
\label{antenna_scenario_conf}
\end{table}

\subsection{Single-panel two-port codebook}
\label{res-single-panel-2-port}
\begin{figure}[!t]
\includegraphics[width=0.7\columnwidth]{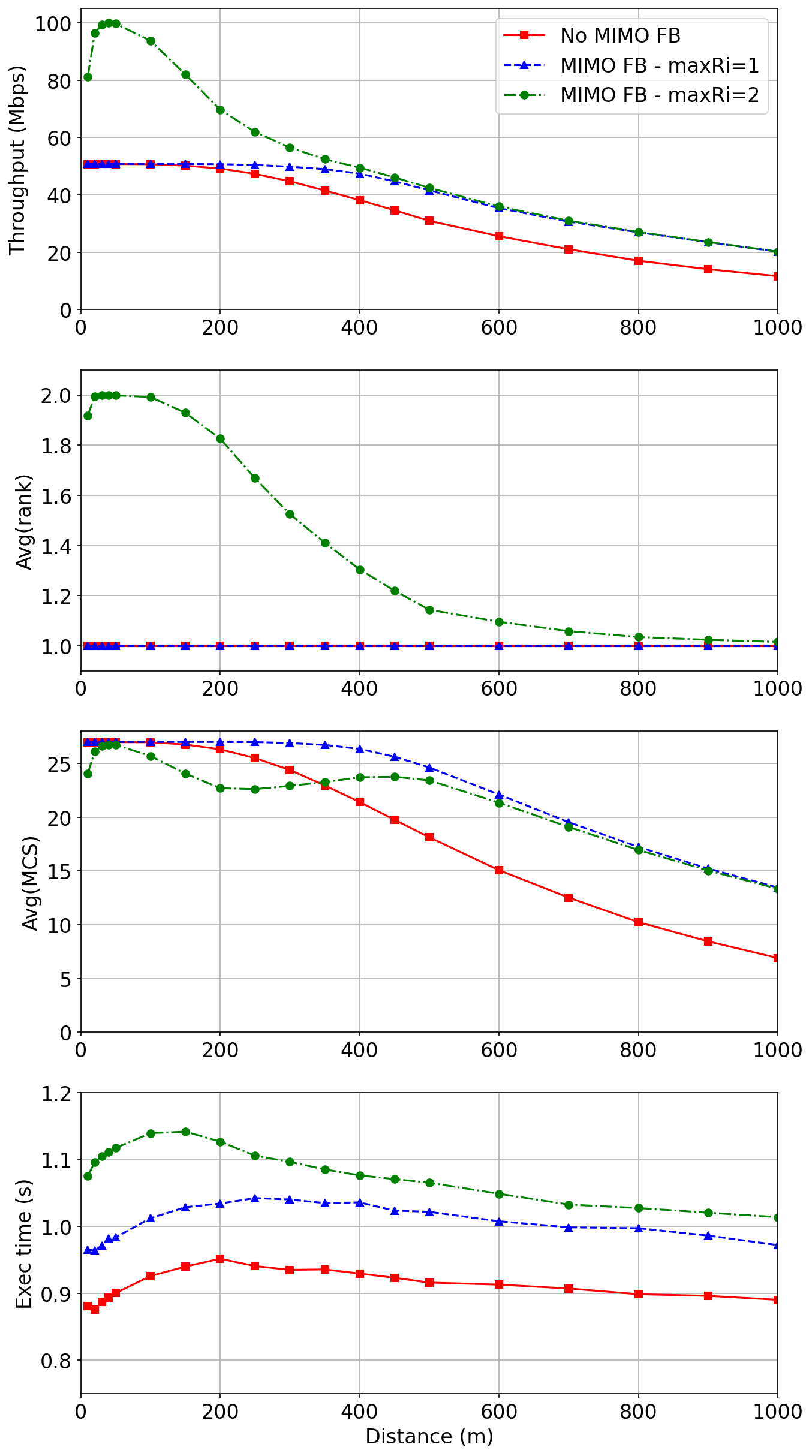}
\centering
\caption{Two-ports codebook (Table 5.2.2.2.1-1)}
\label{conf1}
\end{figure}

The single-panel two-port codebook simulation campaign covers three different cases: 1) when MIMO feedback is not enabled (``No MIMO FB'' in figures), 2) when MIMO feedback is enabled, but the maximum rank is limited to 1 (``MIMO FB - maxRi=1'' in figures), and 3) when MIMO feedback is enabled, and the maximum rank is not limited, i.e., when the maximum rank is set to 2 for two-port antenna, it is not limiting the rank because the maximum rank in such case is 2 (``MIMO FB - maxRi=2'' in figures). In this simulation campaign, the gNB and UE antennas are configured as Conf 1 in Table~\ref{antenna_scenario_conf}. 
In Figure~\ref{conf1}, we show the results of the simulation campaign when using a two-port codebook as defined in TS 38.214 in Table 5.2.2.2.1-1~\cite{TS38214}.

We observe that the MIMO feedback helps achieve better throughput compared to the case when it is disabled. This is thanks to the digital precoding that results in a higher SINR and hence more often is selected higher MCS, than in the case without MIMO feedback. This results in an increased throughput for distances larger than 200 meters. As expected, MIMO with up to 2 streams results in the highest performance, but only for distances up to around 400 meters, where rank 2 is often selected. It can be seen that this happens until around 400 meters, and while the average RI is around 1.3. After, 400 meters it is less and less likely that the rank selected will be 2, but the MCS slightly starts to increase as the distance increases as more often a single stream is selected. We also see a soft drop in the selection of the MCS when the average rank drops below 1.8, which happens because the MCS for two ranks starts to drop, but since still is sometimes being selected rank 1, the MCS for rank 1 is much higher, which results in average MCS that seems quite stagnant from 200 to 400 meters. To summarize, this is because while the rank is dropping at the same time is being more and more selected rank 1 which on average has a higher MCS than the case when rank 2 is selected.

Finally, Figure~\ref{conf1}, also shows the execution time, and as expected, when MIMO feedback is enabled we have a more computationally complex simulation because of the exhaustive PM search explained in Section \ref{pm-search}. Also, when the rank is limited to 1 since the PM search space reduces, the execution time is lower than in the case of maxRi = 2.

\subsection{Single-panel 32-port codebook}
\label{res-single-panel-32-port}
The single panel 32-port codebook simulation campaign tests the correct behavior of the 32-port codebook implemented according to 3GPP TS 38.214~\cite{TS38214}. 
We configure two different simulation campaigns for two different configurations of antenna elements and ports at gNB:
\begin{itemize}
    \item 16 AEs and 8 ports at gNB: the antenna elements at gNB and UE are configured by using the configuration Conf 2a and Conf 2, respectively, from~Table~\ref{antenna_scenario_conf}.
    \item 64 AEs and 16 ports at gNB: the antenna elements at gNB and UE are configured by using the configuration Conf 2b and Conf 2, respectively, from~Table~\ref{antenna_scenario_conf}.
\end{itemize}
In this simulation campaign, we have 5 different cases: MIMO feedback is not enabled (``No MIMO FB'' in figures), and the 4 cases when MIMO feedback is enabled and the rank limit is set to 1, 2, 3, and 4 (i.e., ``MIMO FB - maxRi=1,2,3,4'' in figures). When the rank limit is set to 4, the rank is not limited, because the maximum rank per UE is 4, since as per 3GPP, a single transport block supports the maximum of 4 streams.

\begin{figure}[!t]
\includegraphics[width=0.7\columnwidth]{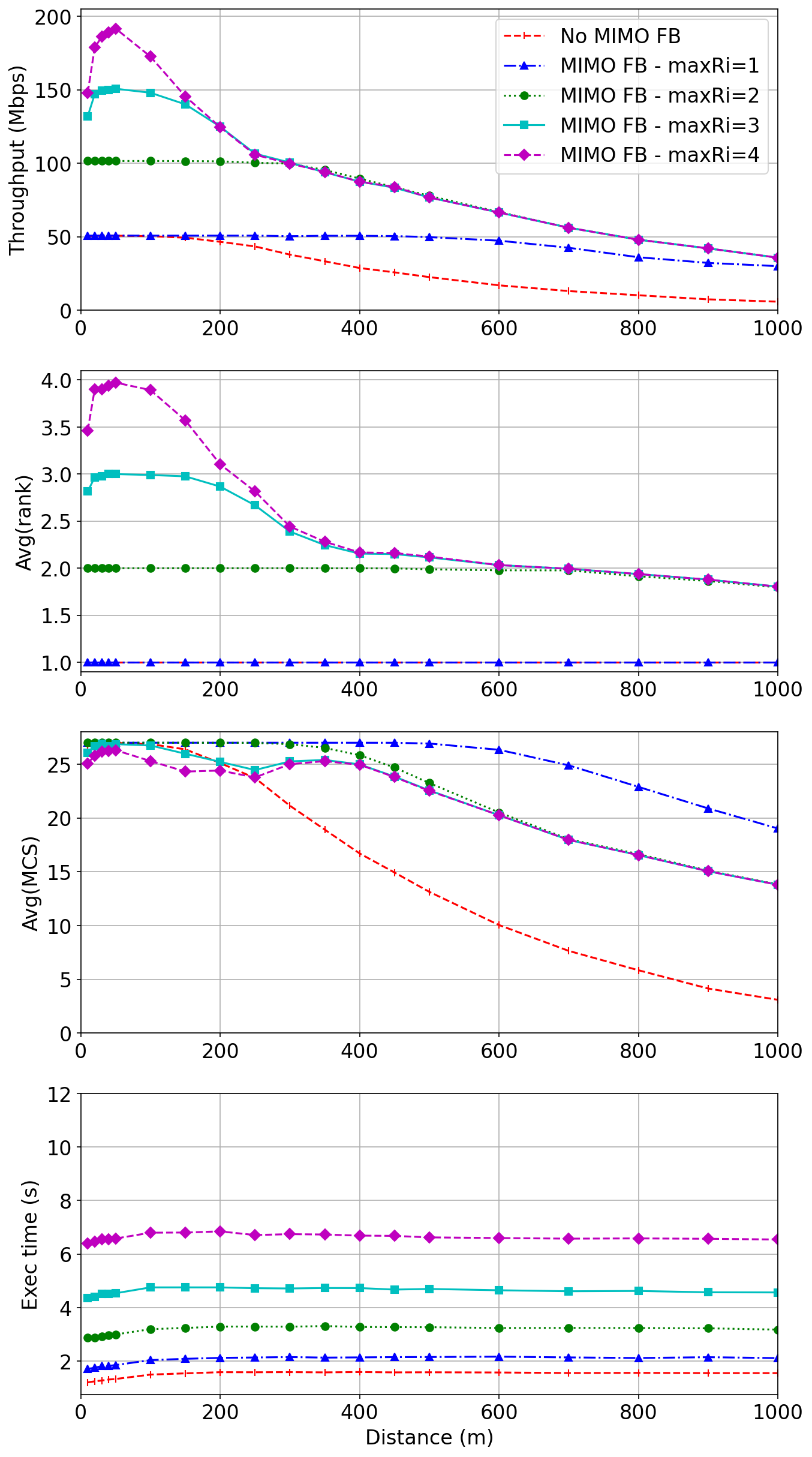}
\centering
\caption{Single-Panel Codebook 3GPP TS 38.214, 16 AEs, 8 ports}
\label{conf2a}
\end{figure}

\begin{figure}[!t]
\includegraphics[width=0.7\columnwidth]{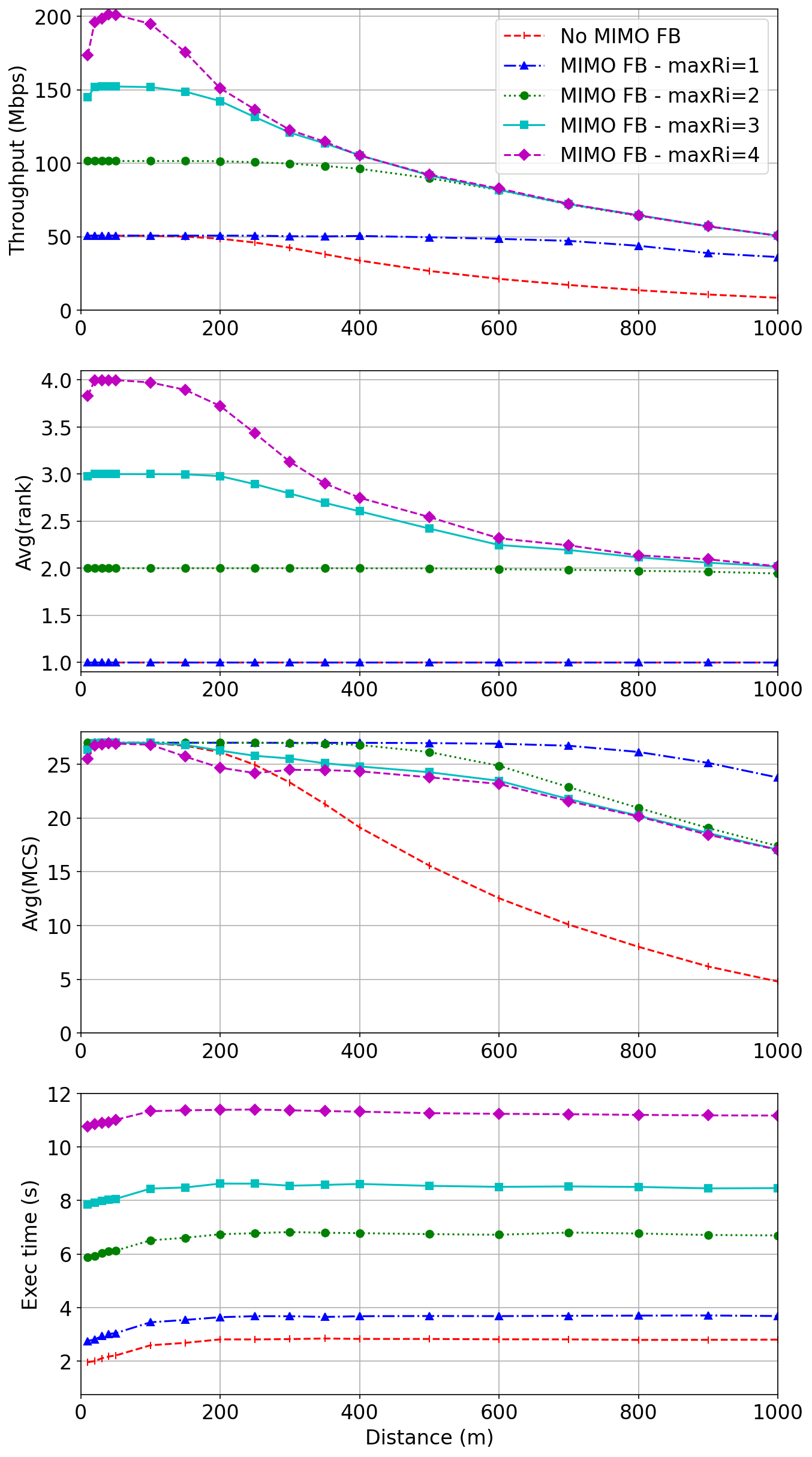}
\centering
\caption{Single-Panel Codebook 3GPP TS 38.214, 64 AEs, 16 ports}
\label{conf2b}
\end{figure}
In Figures \ref{conf2a} and \ref{conf2b} we illustrate the 
performance for these two configurations of the gNB antenna, Conf 2a and Conf 2b, respectively. We observe that in both configurations of the gNB antenna, the SINR is sufficiently large to support rank 2 for almost all simulated distances. The reason for this is also that in this scenario there is no interference, so with large antenna arrays and multiple ports, the signal quality improves significantly. We observe that for lower distances when the rank limit is set to 4, the UE achieves a throughput of around 190~Mbps. The maximum throughput that can be achieved is 200 Mbps, which is 4 times of SISO case which supports at most 50 Mbps, as MCS Table 2 from TS 38.214~\cite{TS38214} is configured. We also observe when increasing the distance how different curves are joining. First maxRi=4 join maxRi=3, at around 200 meters in Conf 2a and around 300 meters in Conf 2b. Then they both join the curve for maxRi=2 at around 300 meters in Conf 2a and around 500 meters in Conf 2b. Finally, these 3 eventually join the maxiRi=1 curve at some larger distance than 800 meters. We see a similar but reverse effect when looking at the average MCS, i.e., the maxRi=4 has the lowest average MCS since it has to split its power into 4 streams, hence its SINR is lower, then this curve joins with the maxRi=3, then later with maxRi=2 and finally with maxRi=1. We observe also a stagnant trend of the average MCS when maxRi is 2, 3, and 4. This is because while the selected rank drops with distance, the average MCS per rank being used increases, which on average gives a similar value for the average MCS for a large range of distances, e.g., for Conf 2a, even if the average rank constantly drops until around 400 meters, we can observe that the MCS is almost constant until the same distance. 

Regarding the computational complexity, we can observe that the complexity increases with the number of antenna elements and the number of ports. 
The large number of antenna elements increases the complexity of the operations with channel matrices. On the other hand, the increased number of ports increases the complexity of the calculation of the spectrum channel matrices, MIMO interference and the PM search. While in Conf 1 the execution time was around 1 second with maxRi=2, we observe for Conf 2a and Conf 2b that the execution time for maxRi=2 has increased 3 and 6 times, respectively. Then further time increases in Conf 2a and 2b increase linearly with maxRi.

\subsection{Impact of antenna ports configuration}
\label{res-antenna-ports}
\begin{figure}[!t]
\includegraphics[width=.7\columnwidth]{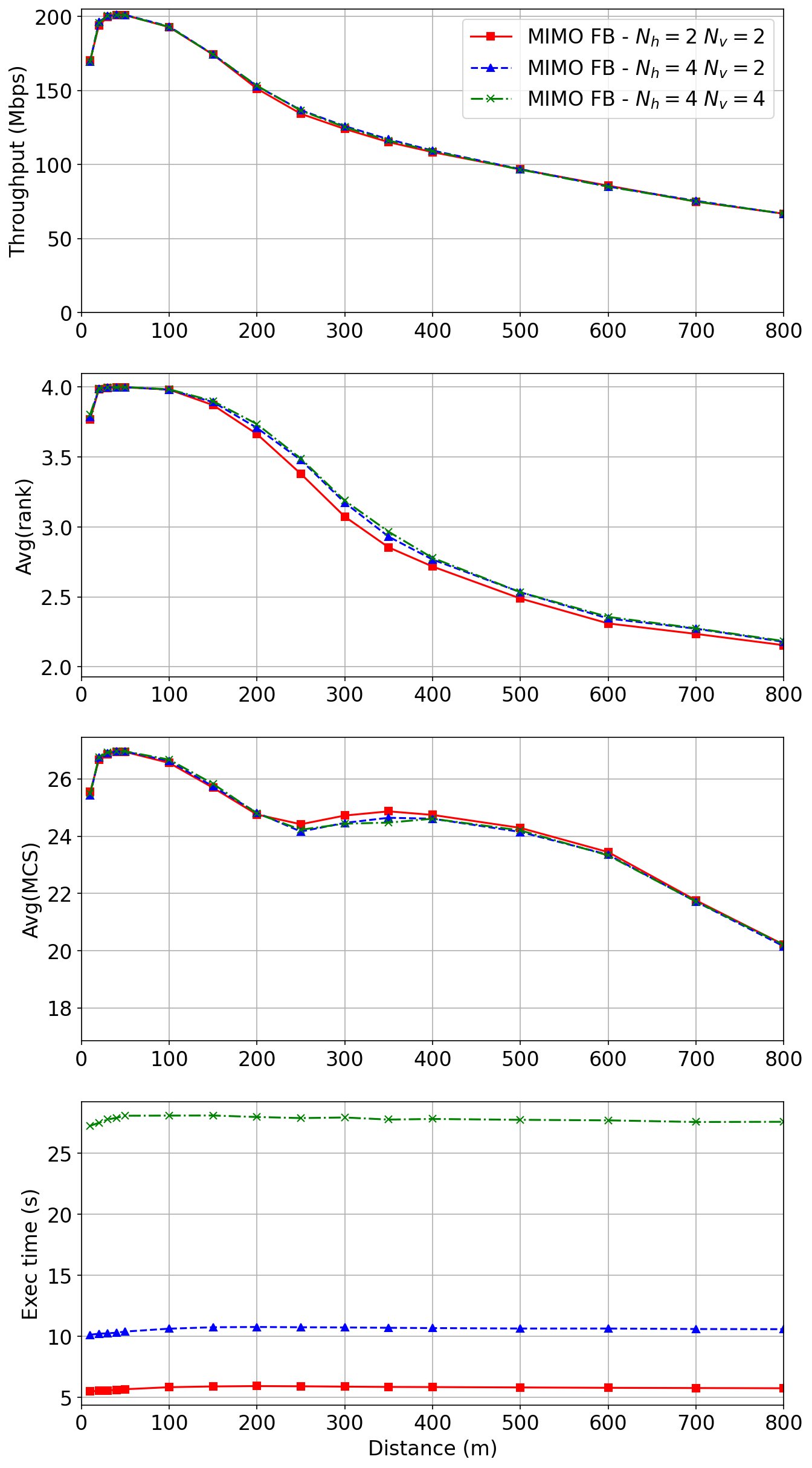}
\centering
\caption{Different gNB ports configurations 8, 16, and 32}
\label{conf3}
\end{figure}

In this section, we analyze the impact of the different port configurations, mainly on the throughput and execution performance, when using an antenna array of a fixed number of dual-polarized antenna elements. Here, 
it is important to remember that the UE moves from gNB in the x-axis direction and that the bearing angles as shown in Table \ref{mimo_scenario_configuration} are configured so that gNB and UE antenna arrays are pointing one to another in the xz plane. We configure 3 different cases of gNB antenna ports:
\begin{enumerate}
    \item 8 ports, $N_{h}=2, N_{v}=2$
    \item 16 ports, $N_{h}=4, N_{v}=2$, and 
    \item 32 ports, $N_{h}=4, N_{v}=4$.
\end{enumerate}
As shown in Table~\ref{antenna_scenario_conf}, in each configuration the gNB antenna array has the same number of antenna elements, which is 64, i.e., 32 in each polarization.
In Figure~\ref{conf3}, we show the performance for different gNB port configurations.

The throughput curves almost overlap, which is the direct result of the overlap of the average rank and the average MCS curves. This confirms that the MIMO implementation provides a correct balance between the analog and the digital beamforming (i.e., when using more transmit antenna ports, we have more gain from the digital beamforming but less from the analog beamforming). It is interesting to see that both analog and digital provide similar gains, and this is why the curves are so similar, because of the near-perfect alignment of gNB and UE antennas. Also, as expected the increase in the vertical ports does not help, instead, the impact of increasing the horizontal ports is slightly noticeable in the average rank and the average MCS curve, which makes sense in mainly 2D deployments. The results of this campaign validate the MIMO implementation. As expected, the execution time increases with the increase in the number of ports. Therefore, in situations with good alignment of antenna arrays, it is better to use analog beamforming and a lower number of antenna ports, to have less complex gNB/UE device operations.

\begin{figure}[!t]
\includegraphics[width=.7\columnwidth]{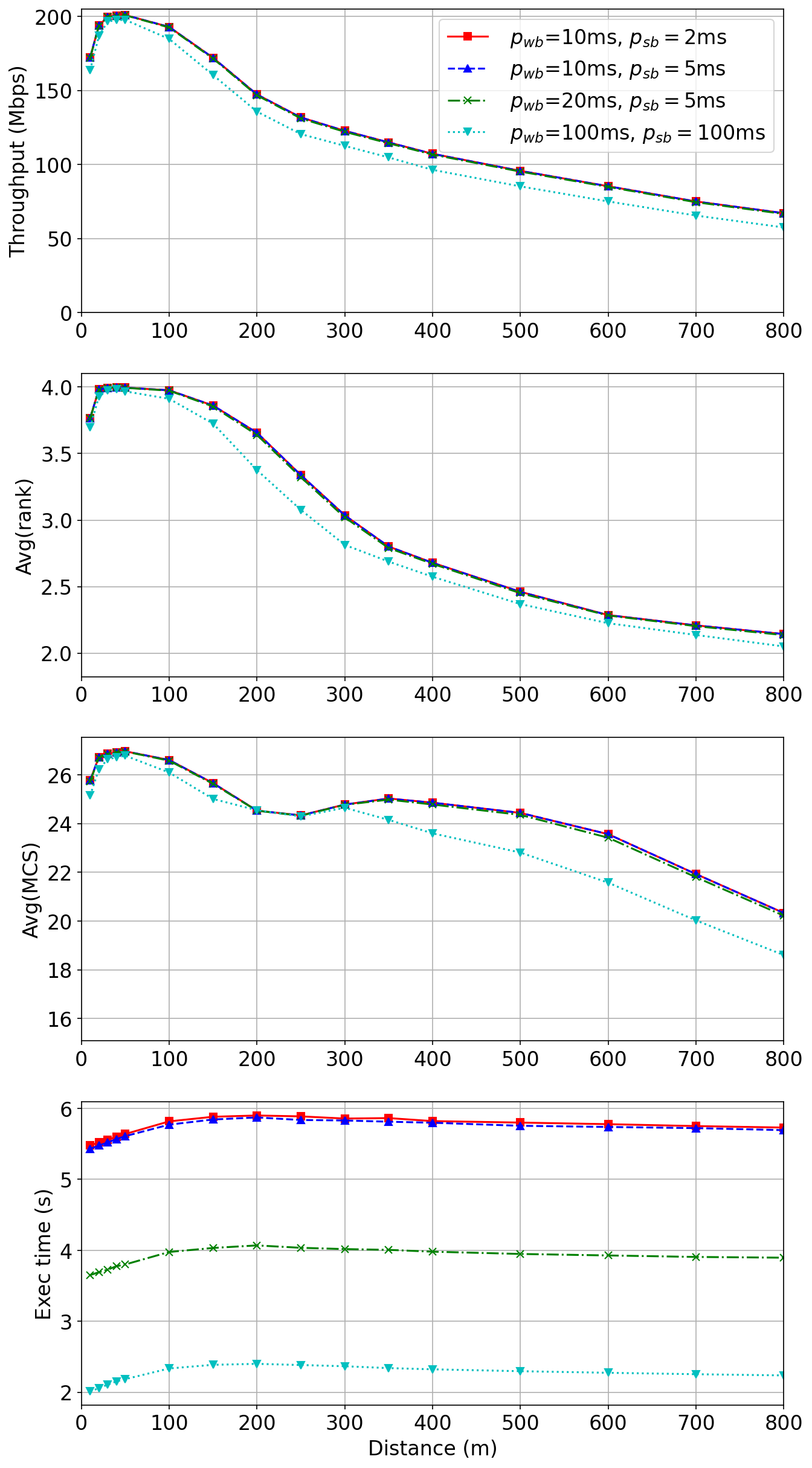}
\centering
\caption{Impact of different SB and WB PMI periodicity}
\label{conf4}
\end{figure}

\subsection{Impact of different SB and WB PMI update intervals}
\label{res-sb-wb-intervals}
Even if in the ideal case the SB and WB precoding matrix should be updated on each received 
signal from the gNB, to reduce the computational complexity, we have 
created two parameters that allow us to configure how often the WB and SB precoding matrices will be updated. This follows the 3GPP standard in the sense that CSI reporting is over a certain period. In this simulation campaign, we vary some typical values that we use in the simulation campaigns, such as 10ms and 2ms for WB PMI and SB PMI updates, respectively. However, we also set a much higher value to analyze the impact that the update frequency has on the final achieved throughput and the computational complexity. 

In Figure~\ref{conf4} we illustrate the impact of different SB and WB PMI periodicities. We observe that there is almost no difference in achieved throughput for the three configurations with the frequencies $p_{wb}=10$ms, $p_{wb}=20$ms, but we can notice an important difference in the simulation execution time between the $p_{wb}=10$ms and $p_{wb}=20$ms, i.e., the simulation time increases around 1.5 times. In the fourth case in which we set $p_{wb}=100$ms, $p_{sb}=100$ms, we observe a drop in the throughput performance of about less than 7.5\%, but we achieve the speedup in the simulation execution of almost 3 times.

\begin{figure}[!t]
\includegraphics[width=.7\columnwidth]{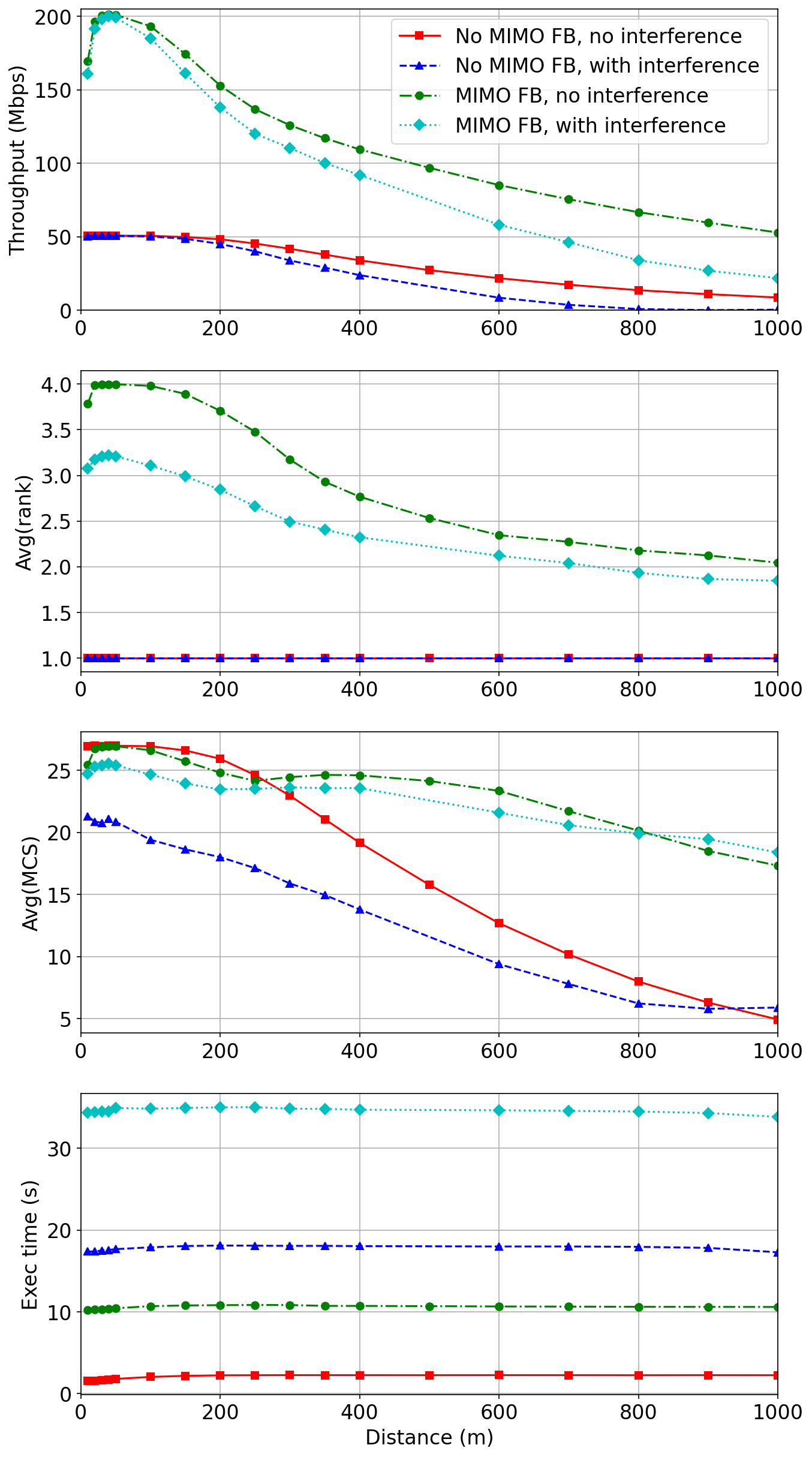}
\centering
\caption{Impact of the interference}
\label{conf5}
\end{figure}

This simulation campaign shows that in the real system if the maximum 
throughput is not needed, the energy could be saved by reducing the frequency of the WB and SB PMI updates. On the other hand, in the simulator, we could significantly reduce the execution time, even without impacting the performance, e.g. reducing the WB PMI update frequency from $10$ms to $20$ms~\footnote{These values are not cross-checked against the 3GPP standard and common cellular configuration parameters, but just simulated to see the impact of various periodicities.}.

\subsection{Scenario with the interference}
\label{eval-interf}
The original topology composed of a single gNB and UE pair is extended to support the possibility of enabling interfering gNB and UE pair. The interfering gNB is placed at a fixed distance of 
1000 meters from the original gNB and it is placed on the x-axis its position is (x,y,z) = (1000,0,25). The UE of the interfering gNB and UE pair is placed at the fixed position with coordinates (x,y,z)=(500, 0, 1.5). The bearing angles of the interfering gNB and UE are 180 and 0, respectively. 
The antennas of the original and interfering gNB are configured according to Conf 3 with 16 ports, while UEs' antennas are configured as in Conf 2. In Figure~\ref{conf5} we illustrate the performance of the four cases: when the 
MIMO feedback is disabled and enabled and when the 
interfering node is not present and when it is. 
As expected, we observe better throughput performance when the interfering node is not enabled regardless of whether the MIMO feedback is enabled. 
When MIMO feedback and the interfering node are enabled, we observe that on average a lower rank is selected, leading to very similar average values of MCS, i.e., when the interfering node pair is present the lower rank is on average selected, which allows using on average a higher MCS. In this simulation 
campaign, we can observe a huge increase in the execution time when the interfering node pair is enabled. Also, the simulation time increases by more than 15x when having just an additional node pair and the MIMO FB is enabled.

\section{Considerations on MIMO models fidelity vs computational complexity in ns-3 network simulator}\label{sec:tradeoff}

ns-3 and 5G-LENA simulators are network simulators, as such their main objective is to allow full-stack end-to-end and large-scale network simulations. 
In addition, NGMN and 3GPP extended reality and cloud gaming traffic models have been recently included in the 5G-LENA simulator, to allow the simulation of different ``hot topic" scenarios related to augmented, virtual, and mixed reality in 5G and beyond~\cite{10120945}.
As mentioned earlier, the academy and industry are each time more and more interested in having a high-fidelity PHY layer in network simulators such as ns-3 and 5G-LENA. This poses many challenges related to the dramatically increased computational complexity of such simulations. As it could be observed through multiple simulation campaigns presented in previous sections, the MIMO model adds a level of computational complexity to the already computationally expensive ns-3 3GPP spatial channel model. 
One of the main bottlenecks is the exhaustive PM search method, which is executed frequently when MIMO is enabled. For this reason, we started recently to work on alternative solutions for PM search that would cut down the simulation execution time. We consider this as a very interesting research topic, for two reasons: 1) the ns-3 and 5G-LENA simulators need much lighter PHY models to allow its main promise which is to perform large-scale end-to-end simulations, and 2) improvements in the PM search algorithm are of crucial interest for the real UE devices, e.g., mobile phones, where the computational complexity translates directly into energy consumption. Hence, to allow the use of MIMO in real devices without draining a UE device's battery, it is necessary to come up with computationally efficient PM search algorithms, which at the same time provide good performance in selecting the rank and PMI.

Yet another performance degradation that was introduced with the MIMO model is due to the removal of the ``OFDMA downlink trick". For example, in the scenario in which we have a single gNB and 10 UEs, in the case of OFDMA transmissions, if gNB was transmitting at the same time to 10 UEs, we would have only a single transmit signal entering into the spectrum channel. Since 10 UEs are listening, that means that we had 10 times called functions \texttt{DoCalcRxPowerSpectralDensity}. However, now instead of 1 signal, we have an independent signal per each UE, hence, 10 transmit signals will enter into the spectrum channel, and each of them will be received by 10 UEs, resulting in 100 calls of \texttt{DoCalcRxPowerSpectralDensity}. This causes the degradation of the 5G-LENA simulator performance, even when MIMO is disabled. We plan to solve this soon, as soon as CSI-RS is implemented and hence each UE in the OFDMA system will have a reference signal based on which it will be able to estimate the channel in all the RBs. Meanwhile, there is no CSI-RS implementation, the 5G-LENA model needs each UE to receive all these signals that are not intended for that device, to be able to estimate the channel over all the RBs and perform the PM search in the case of MIMO. Once CSI-RS is implemented, we can filter out undesired signals (i.e., discarding by RNTI identification or based on the non-overlapping RBs).



\section{Conclusions and Future work}
\label{sec:conc}
In this work, we proposed, implemented, and evaluated a full SU-MIMO model for system-level simulation of 5G NR MIMO systems in ns-3, by extending the current framework with the support of MIMO spatial multiplexing to send multiple data streams towards the same user, thus enabling the hybrid beamforming feature of 3GPP. 

The code is publicly available and merged in the master branch of ns-3 5G-LENA module~\cite{5gLenaRelease3}. In this paper, we focused mainly on the most novel aspects of this MIMO implementation in the ns-3 simulator and that is the MIMO digital baseband processing. We have implemented the codebook Type-I precoding (for single panel case), assuming MMSE-IRC receiver for inter-stream interference suppression, as adopted in 3GPP. 
Compared to the old DP-MIMO implementation, the new model provides a more generic and realistic model for DP-MIMO in which the streams, ports, and antenna mapping are not predefined, and instead are defined through digital precoding. Thanks to this feature a more accurate model for the
inter-stream interference calculation is considered. The current model supports up to 32 antenna ports at gNB and UE and 4 ranks per user. 

Through multiple simulation campaigns, we tested different aspects of the newly implemented full SU-MIMO model, and the results showed that the model performs correctly. New MIMO interference computations and PM search algorithms increase significantly the computational complexity of the 5G-LENA simulator. In our future work, we plan to implement less computationally expensive PM search algorithms and also to provide multiple optimizations that would reduce the computational complexity of the MIMO interference calculations. Such improvements are essential for the use of the implemented MIMO models in large-scale network simulation scenarios. We also plan to extend the current MIMO model to support uplink MIMO and MU-MIMO through the support of Precoding Type-II and new MU-MIMO scheduling mechanisms.

\section*{Acknowledgment}
The work has received funding from Grant PID2021-126431OB-I00 funded by MCIN/AEI/10.13039/501100011033 and “ERDF A way of making Europe”, TSI-063000-2021-56/57 6G-BLUR project by the Spanish Government, and Generalitat de Catalunya grant 2021 SGR 00770.

\bibliography{main}

\end{document}